\documentclass[twocolumn,english,amsart,showpacs,preprintnumbers,amsmath,amssymb,floatfix]{revtex4-2}
\usepackage{tikz,xcolor}
\usepackage[colorlinks = true,
linkcolor = blue,
urlcolor  = blue,
citecolor = blue,
anchorcolor = blue]{hyperref}

\definecolor{lime}{HTML}{A6CE39}
\DeclareRobustCommand{\orcidicon}{%
	\begin{tikzpicture}
	\draw[lime, fill=lime] (0,0)
	circle [radius=0.16]
	node[white] {{\fontfamily{qag}\selectfont \tiny ID}};
	\draw[white, fill=white] (-0.0625,0.095)
	circle [radius=0.007];
	\end{tikzpicture}
	\hspace{-2mm}
}

\foreach \x in {A, ..., Z}{%
\expandafter\xdef\csname orcid\x\endcsname{\noexpand\href{https://orcid.org/\csname orcidauthor\x\endcsname}{\noexpand\orcidicon}}
}

\usepackage[T1]{fontenc}
\usepackage[latin9]{inputenc}
\usepackage{color}
\usepackage{array}
\usepackage{amstext}
\usepackage{graphicx}
\usepackage{esint}
\usepackage{rotating}
\usepackage{appendix}
\usepackage{float}

\usepackage[font=small,labelfont=bf]{caption}
\usepackage{xcolor}
\usepackage{ulem}

\makeatletter

\providecommand{\tabularnewline}{\\}

\@ifundefined{textcolor}{}
{%
 \definecolor{BLACK}{gray}{0}
 \definecolor{WHITE}{gray}{1}
 \definecolor{RED}{rgb}{1,0,0}
 \definecolor{GREEN}{rgb}{0,1,0}
 \definecolor{BLUE}{rgb}{0,0,1}
 \definecolor{CYAN}{cmyk}{1,0,0,0}
 \definecolor{MAGENTA}{cmyk}{0,1,0,0}
 \definecolor{YELLOW}{cmyk}{0,0,1,0}
 }

\@ifundefined{definecolor}
 {\usepackage{color}}{}
\@ifundefined{definecolor}
 {\usepackage{color}}{}
\makeatother
\usepackage{babel}

\begin{document}

%%%%%%%%%%%%%%%%%%%%%%%%%%%%%%

\title{Recent data analysis to revisit the spin structure function of nucleon in Laplace space}

\author{Hoda Nematollahi$^{1}$\orcidA{}}
\email{hnematollahi@uk.ac.ir }

	\author{ Abolfazl Mirjalili$^{2}$\orcidC{}}
	\email{A.Mirjalili@yazd.ac.ir}

\author {Shahin Atashbar Tehrani$^{3}$\orcidD{}}
\email{Atashbar@ipm.ir}

\affiliation {
	$^{(1)}$Faculty of Physics, Shahid Bahonar University of Kerman, Kerman, Iran\\
     $^{(2)}$Physics Department, Yazd University, Yazd, Iran\\
	$^{(3)}$School of Particles and Accelerators, Institute for Research in Fundamental Sciences (IPM), P.O.Box
	19395-5531, Tehran, Iran}

\date{\today}

\begin{abstract}\label{abstract}
Considering a fixed-flavor number scheme and based on laplace transdormation,  we perform a leading-order and next-to-leading-order QCD analysis which are including world data on polarized structure functions $g_1$ and $g_2$. During our analysis, taking the DGLAP evolution,  we employ the  Jacobi polynomials expansion technique. In our recent analysis we utilize the recent available data and consequently include more data than what we did in our previous analysis. we obtain good agreements between our results for the polarized parton densities and nucleon structure functions with all available experimental data and some common parametrization models.
\end{abstract}

\maketitle
\tableofcontents{}

%%%%%%%%%%%%%%%%%%%%%%%%%%%%%%%%%%%%%%%%%%    Introduction    %%%%%%%%%%%%%%%%%%%%%%%%%%%%%%%%%%%%%%%%%%%%%%%%%
\section{Introduction}\label{sec:sec1}
High-energy scattering of polarized leptons by polarized protons, neutrons, and deuterons provides a measurement of the nucleon spin structure functions. These structure functions give information on the polarized quark contributions to the spin of the proton and the neutron and allow tests of the quark-parton model and quantum chromodynamics (QCD).
After precise consideration of the unpolarized deep inelastic scattering (DIS) experiments, polarized DIS program has been planed to study the spin structure of the nucleon using polarized lepton beams (electrons and muons) scattered by polarized targets.
These fixed-target experiments have been used to characterize the spin structure of the proton and
neutron and also to test fundamental sum rules of QCD and quark-parton model (QPM) \cite{nsp1,DBT2019}. The first experiments in polarized electron-polarized proton scattering, performed around 50 years ago, helped to establish the parton structure of the proton.
About two decades later, by performing an experiment with polarized muon and polarized proton, this reality has been raveled that the QPM sum rule was violated which seemed to indicate that the quarks do not contribute alone to the spin of the proton.
This ``proton-spin crisis'' gave birth to a new generation of experiments at several high-energy physics laboratories
around the world. The new and extensive data sample, collected from these fixed target experiments, has enabled a careful characterization of the spin-dependent parton substructure of the nucleon. The results have been used to test QCD, to find
an independent value for $\alpha_s(Q^2)$, to probe the polarized parton distributions with reasonable precision, and to provide a first look at the polarized gluon distribution \cite{nsp2}.

On this purpose we try to solve the Dokshitzer-Gribov-Lipatov-Altarelli-Parisi (DGLAP) evolution equations, using the Laplace transformation. This is done at leading order (LO) and next-to-leading order (NLO) approximations in SubSec.\ref{sec:sec2A} and \ref{sec:sec2B} of Sec.\ref{sec:sec2}. We then construct the $xg_1$ polarized structure function using the expansion in terms of Jaccobi polynomials in  Sec.\ref{sec:4}. The evolution of partons requires some inputs which are in fact the polarized parton distribution functions (PPDFs) at initial energy scale, $Q_0^2$. On this base we need to some parameterizations for the input PPDFs which is introduced in Sec.\ref{sec:5}. To determine the unknown parameters of the input PPDFs we should take all the recent and available data from different DIS experiments. We then use them in a fitting process as is illustrated in Sec.\ref{sec666}. Getting the $g_1$ structure function, it is possible to calculate the $g_2$ structure function which is done also in this section. To validate the results from  data analysis for $g_1$ structure function, several sum rules are computed. We find them  in good agreement with experimental data and the results which are arising out from theoretical investigations. This part of work is presented in Sec.\ref{sec:sec7}. The worthiness of the work which we handle is to employ the last reported data for polarized targets in DIS experiments. The compatible results with data at different energy scales and some models confirm authenticity of the utilized theoretical framework, including QPM and some outputs of QCD which are described in Sec.\ref{sec:sec8} as the last section of this paper.
\section{Polarized DGLAP evolution equations in Laplace space}\label{sec:sec2}
\subsection{Leading-order approximation}\label{sec:sec2A}
In this work we generalize the method of  Laplace transformation to employ QCD evolution equations in polarized case to investigate the polarized parton distributions of the nucleon. Here we focus on the polarization of singlet and non-singlet quarks to indicate the efficiency of this method for solving the DGLAP evolution equations \cite{evol1,evol2,evol3,evol4}. In order to extract the polarized parton distribution functions, we review the method in this section briefly.

By introducing the variable $\nu\equiv ln(\frac{1}{x})$ into leading order coupled DGLAP equations, it is possible to turn them into coupled convolution equations in $\nu$ space. One can use two Laplace transformations, one from $\nu$ space to $s$ space and the second one from $\tau$ space to $U$ space, using new variable $%
 \tau \equiv \frac{1}{4\pi }\int_{Q_{0}^{2}}^{Q^{2}}\alpha
 _{s}(Q^{\prime 2})d\ln Q^{\prime 2}$. The DGLAP evolution equations can be solved using these two laplace transformations and set of convolution integrals of polarized parton distributions at the initial scale $Q_0^2$. Finally by applying two inverse laplace transforms, we can back to ordinary space (x, $Q^2$) \cite{Block:2010du}.

The DGLAP evolution equations for the polarized parton distributions are written as \cite{evol1,evol2,evol3,evol4,Furmanski:1981cw}:
 \begin{equation}
 \frac{4\pi }{\alpha _{s}(Q^{2})}\frac{\partial \Delta
 F_{NS}}{\partial \ln Q^{2}}(x,Q^{2})=\Delta F_{NS}\otimes \Delta
 P_{qq}^{0}(x,Q^{2})\;,
 \label{eq:1}
 \end{equation}

 \begin{equation}
 \frac{4\pi }{\alpha _{s}(Q^{2})}\frac{\partial \Delta
 	F_{S}}{\partial \ln Q^{2}}(x,Q^{2})=\Delta F_{S}\otimes \Delta
 P_{qq}^{0}+\Delta G\otimes \Delta P_{qg}^{0}(x,Q^{2})\;,
 \end{equation}

 \begin{equation}
 \frac{4\pi }{\alpha _{s}(Q^{2})}\frac{\partial \Delta G}{\partial \ln Q^{2}}%
 (x,Q^{2})=\Delta F_{S}\otimes \Delta P_{gq}^{0}+\Delta
 G\otimes \Delta P_{gg}^{0}(x,Q^{2})\;,
 \end{equation}
where $\Delta P_{ij}^{0}$'s are the LO polarized splitting functions.

By introducing the variable change $w\equiv \ln (1/z)$, applying two other ones: $v\equiv \ln (1/x)$ and $%
 \tau (Q_0^2,Q^2) \equiv \frac{1}{4\pi
 }\int_{Q_{0}^{2}}^{Q^{2}}\alpha _{s}(Q^{\prime 2})d\ln Q^{\prime
 2}$, introduced before, and using the notation $\Delta \widehat{F}_{NS}(\nu,\tau )\equiv
\Delta F_{NS}(e^{-v},Q^{2})$ , $\Delta \widehat{F}_{NS}(w,\tau
)\equiv \Delta F_{NS}(e^{-w},\tau )$, $\Delta
\widehat{F}_{s}(\nu,\tau )\equiv \Delta F_{s}(e^{-v},Q^{2})$,
$\Delta \widehat{F}_{s}(w,\tau )\equiv \Delta F_{s}(e^{-w},\tau
)$, $\Delta \widehat{G}(v,\tau )\equiv \Delta G(e^{-v},\tau)$, $\Delta \widehat{G}%
_{s}(w,\tau )\equiv \Delta G_{s}(e^{-w},\tau )$, the above DGLAP equations in terms of the convolution integrals are given as:

\begin{equation}
\frac{\partial \Delta \widehat{F}_{NS}}{\partial \tau }(v,\tau
)=\int_{0}^{v}\Delta \widehat{F}_{NS}(w,\tau )\Delta
\widehat{H}_{qq}^{0}(v-w)dw\;,\label{eq:ns}
\end{equation}

\begin{eqnarray}
&&\frac{\partial \Delta \widehat{F}_{S}}{\partial \tau }(v,\tau
)=\int_{0}^{v}\Delta \widehat{F}_{s}(w,\tau )\Delta
\widehat{H}_{qq}^{0}(v-w)dw +\nonumber\\
&&\int_{0}^{v}\Delta
\widehat{G}_{s}(w,\tau )\Delta \widehat{H}_{qg}^{0}(v-w)dw\;,\nonumber\\
\label{eq:fs}
\end{eqnarray}

\begin{eqnarray}
&&\frac{\partial \Delta \widehat{G}}{\partial \tau }(v,\tau
)=\int_{0}^{v}\Delta \widehat{F}_{s}(w,\tau )\Delta
\widehat{H}_{gq}^{0}(v-w)+\nonumber\\
&&\int_{0}^{v}\Delta
\widehat{G}_{s}(w,\tau )\Delta \widehat{H}_{gg}^{0}(v-w)dw\;,\nonumber\\
\label{eq:g}
\end{eqnarray}
where
\begin{eqnarray}
\Delta \widehat{H}_{qq}^{0}(v) &\equiv &e^{-v}\Delta P_{qq}^{0}(e^{-v}), \nonumber\\
\Delta \widehat{H}_{gq}^{0}(v) &\equiv &e^{-v}\Delta
P_{gq}^{0}(e^{-v}),  \nonumber
\\
\Delta \widehat{H}_{qg}^{0}(v) &\equiv &e^{-v}\Delta
P_{qg}^{0}(e^{-v}),  \nonumber
\\
\Delta \widehat{H}_{gg}^{0}(v) &\equiv &e^{-v}\Delta
P_{gg}^{0}(e^{-v})\;.
\end{eqnarray}
By considering the following property of laplace transforms:
\begin{eqnarray}
{\cal L}\left[ \int_{0}^{v}\Delta \widehat{F}[w]\Delta \widehat{H}[v-w%
]dw;s\right]\nonumber\\
={\cal L}[\Delta \widehat{F}_{s}[v];s] \times {\cal
	L}\left[ \Delta \widehat{H}[v];s\right],
\end{eqnarray}
the DGLAP equations in Eqs.(\ref{eq:ns}, \ref{eq:fs}) and Eq.(\ref{eq:g}) can be converted to three coupled ordinary first-order differential equations in terms of the variable $\tau$ in the Laplace s-space with $\tau$-dependent coefficients as following:
\begin{eqnarray}
\frac{\partial \Delta f_{NS}}{\partial \tau }(s,\tau ) &=&\Delta
\Phi _{ns}^{LO}(s)\Delta f(s,\tau )\;,\nonumber \\
\frac{\partial \Delta f_{s}}{\partial \tau }(s,\tau ) &=&\Delta
\Phi _{f}^{LO}(s)\Delta f_{s}(s,\tau )+\Delta \Theta
_{f}^{LO}(s) \Delta g(s,\tau)\;,\label{LO-f}
\nonumber \\
\frac{\partial \Delta g}{\partial \tau }(s,\tau ) &=&\Delta \Phi
_{g}^{LO}(s)\Delta g(s,\tau )+\Delta \Theta
_{g}^{LO}(s)\Delta f_{s}(s,\tau )\;. \nonumber\\
\end{eqnarray}
In the above equations, we utilize the following abbreviations:

\begin{eqnarray}
\Delta f_{NS}(s,\tau )&\equiv& {\cal L}[\Delta
\widehat{F}_{NS}(v,\tau );s],\nonumber\\
\Delta f_{s}(s,\tau )&\equiv& {\cal
	L}[\Delta \widehat{F}_{s}(v,\tau );s],\nonumber\\
\Delta g(s,\tau )&\equiv& {\cal
	L}[\Delta \widehat{G}(v,\tau );s]\;.
\end{eqnarray}
Following that we can write:
\begin{eqnarray}
{\cal L}\left[ \frac{\partial \Delta \widehat{F}_{NS}}{\partial
	w}(w,\tau
);s\right]  &=&s\Delta f_{NS}(s,\tau )\;, \\
{\cal L}\left[ \frac{\partial \Delta \widehat{F}_{s}}{\partial
	w}(w,\tau );s\right]  &=&s\Delta f_{s}(s,\tau )\;, \nonumber\\
{\cal L}\left[ \frac{\partial \Delta \widehat{G}}{\partial
	w}(w,\tau );s\right]  &=&s\Delta g(s,\tau )\;. \nonumber
\end{eqnarray}
On the other hand the LO coefficients  $\Delta \Phi ^{LO}$ and $\Delta \Theta ^{LO}$ in Laplace s-space
are given by:

\begin{equation}
\Delta \Phi _{f}^{LO}=4-\frac{8}{3}\left(
\frac{1}{s+1}+\frac{1}{s+2}+2(\psi^{(0)} (s+1)+\gamma _{E})\right) \label{eq:qq},
\end{equation}

\begin{equation}
\Delta \Theta _{f}^{LO}=T_r\left(
\frac{2}{s+2}-\frac{1}{s+1}\right)\;,
\label{eq:qq1}
\end{equation}

\begin{equation}
\Delta \Theta _{g}^{LO}=C_{f}\left(
\frac{2}{s+1}-\frac{1}{2+s}\right)\;,
\label{eq:gg1}
\end{equation}

\begin{eqnarray}
\Delta \Phi _{g}^{LO}=C_{a}\left( \frac{11}{6}-\frac{f}{9}+\frac{3}{s+1}-%
\frac{3}{s+2}+\frac{1}{s+3}+\right. \nonumber \\ \left. \frac{1}{s+4}+\psi^{(0)} (s+1)-\psi^{(0)}\;.
(s+5)\right)\;.\label{eq:gg}\nonumber\\
\end{eqnarray}
In Eq.(~\ref{eq:qq})and Eq.(~\ref{eq:gg}) $\psi^{(0)}(x)$ denotes digamma function and $\gamma_E=0.5772156$ is Euler's constant.
The evolution of DGLAP equations in Laplace space at the LO approximation for singlet sector and gluon part can be written as:
\begin{eqnarray}
\Delta f_1(s,\tau)=\Delta k_{ff_1}(s,\tau)f_0(s)+\Delta k_{fg_1}(s,\tau)g_0(s)\;,\nonumber\\
\Delta g_1(s,\tau)=\Delta k_{gg_1}(s,\tau)g_0(s)+\Delta k_{gf_1}(s,\tau)f_0(s),
\label{eq:24}
\end{eqnarray}
where the $\Delta k$'s in above equations are given by:
\begin{eqnarray}
\Delta k_{ff_1}(s,\tau) && \equiv e^{\frac{\tau}{2}(\Delta\Phi_f^{LO}(s)+\Delta\Phi_g^{LO}(s))}\left[cosh\left(\frac{\tau}{2}R(s)\right)\right.\nonumber\\
&& \left. +\frac{2sinh(\frac{\tau}{2}R(s))}{R(s)} (\Phi_f^{LO}(s)-\Phi_g^{LO}(s))\right] ,\nonumber\\
\Delta k_{fg_1}(s,\tau) && \equiv e^{\frac{\tau}{2}(\Delta\Phi_f^{LO}(s)+\Delta\Phi_g^{LO}(s))}\frac{sinh\left(\frac{\tau}{2}R(s)\right)}{R(s)}\Delta \Theta_f^{LO}(s),\nonumber\\
\Delta k_{gg_1}(s,\tau) && \equiv e^{\frac{\tau}{2}(\Delta\Phi_f^{LO}(s)+\Delta\Phi_g^{LO}(s))}\left[cosh\left(\frac{\tau}{2}R(s)\right)\right.\nonumber\\
&& \left. -\frac{2sinh(\frac{\tau}{2}R(s))}{R(s)} (\Phi_f^{LO}(s)-\Phi_g^{LO}(s))\right] ,\nonumber\\
\Delta k_{fg_1}(s,\tau) && \equiv e^{\frac{\tau}{2}(\Delta\Phi_f^{LO}(s)+\Delta\Phi_g^{LO}(s))}\frac{sinh\left(\frac{\tau}{2}R(s)\right)}{R(s)}\Delta \Theta_g^{LO}(s).\nonumber\\
\label{eq:251}
\end{eqnarray}
In Eq.(\ref{eq:251}), $R(s)$ is defined as:
\begin{equation}
R(s)=\sqrt{(\Delta\Phi_f^{LO}(s)-\Delta\Phi_g^{LO}(s))^2+4\Delta\Theta_f^{LO}(s)\Delta\Theta_g^{LO}(s)}\;,
\label{eq:23n}
\end{equation}

 For doing the numerical Laplace inversion in $v$ space, one need the Laplace inverse of the kernels
as $K_{FF}(v,\tau)\equiv{\cal L}^{-1}[\Delta k_{ff}(s,\tau);v]$, $K_{FG}(v,\tau)\equiv{\cal L}^{-1}[\Delta k_{fg}(s,\tau);v]$,$\;\;\;$ $K_{GF}(v,\tau)\equiv{\cal L}^{-1}[\Delta k_{gf}(s,\tau);v]$ and $K_{GG}(v,\tau)\equiv{\cal L}^{-1}[\Delta k_{gg}(s,\tau);v]$ ~\cite{Block:2010ti,Block:2010du,Block:2009en}. So that we can write the decoupled solutions in $(v,Q^2)$ space, based on the following convolutions:
\begin{eqnarray}
\Delta\widehat{F}_s(v,Q^2)&&\equiv\int_0^v K_{FF}(v-w,\tau(Q^2,Q_0^2))\Delta\widehat{F}_{so}(w)dw\nonumber\\
&&+\int^v_0  K_{FG}(v-w,\tau(Q^2,Q_0^2))\Delta\widehat{G}_{so}(w)dw\;,\nonumber\\
\Delta\widehat{G}_s(v,Q^2)&&\equiv\int_0^v K_{GG}(v-w,\tau(Q^2,Q_0^2))\Delta\widehat{G}_{so}(w)dw\nonumber\\
&&+\int^v_0  K_{GF}(v-w,\tau(Q^2,Q_0^2))\Delta\widehat{F}_{so}(w)dw;.\nonumber\\
\label{eq:26}
\end{eqnarray}
It is obvious that $\Delta\widehat{F}_{so}(w)$ and $\Delta\widehat{G}_{so}(w)$ are the Laplace inverse of $f_0(s)$ and $g_0(s)$ in Eq.(\ref{eq:24}). Reminding $w\equiv ln(1/z)$  and recalling that $v\equiv ln(1/x)$, we can finally convert the above solutions into the usual Bjorken-$x$ space.\\

Now for non-singlet sector, $\Delta F_{ns}(x,Q^2)$, as before, using the variable change  $v\equiv ln(1/x)$
and the variable $\tau$ then the valance part in Eq.(~\ref{eq:1}) can be written as:
\begin{equation}
\frac{\partial\Delta \widehat{F}_{NS}}{\partial \tau}=\int_0^v \Delta\widehat{F}_{NS}(w,\tau)e^{-(v-w)}\Delta P_{qq}^{LO,ns}(v-w)dw\;.
\label{eq:27}
\end{equation}
Employing the Laplace transformation on above equation, we obtain a linear differential equation in terms of $\tau$ variable  for the $\Delta f_{ns}(s,\tau)$ as the transformed version of $\Delta \widehat{F}_{NS}(x,Q^2)$. This differential equation leads to the following solution:
\begin{equation}
\Delta f_{ns}(s,\tau)\equiv e^{\tau\Delta\Phi_{ns}^{LO}}\Delta f_{ns0}(s)\;.
\label{eq:28}
\end{equation}
Now using the inverse Laplace transform on Eq.(\ref{eq:28}) we arrive at the following convolution:
\begin{equation}
\Delta\widehat{F}_{ns}(v,\tau)=\int_0^v  K_{ns}(v-w,\tau)\Delta\widehat{F}_{ns0}(w)dw\;.
\label{eq:29}
\end{equation}
In this equation we take $\Delta K_{ns}(v,\tau)\equiv{\cal L}^{-1}[e^{\tau \Delta\Phi_{ns}^{LO}(s)};v]$ where $\Delta\widehat{F}_{ns0}(w)$ is the inverse Laplace of $\Delta f_{ns0}(s)$. Finally by variable change $\nu=ln(\frac{1}{x})$ the results in  $(x,Q^2)$ space is accessible. \\

More details to extract parton distribution functions at the LO approximation, based on the Laplace transformation, can be found in \cite{Block:2010du}.
\subsection{Next-Leading-order approximation}\label{sec:sec2B}
At the NLO approximation for the non-singlet sector of DGLAP evolution equation, after changing the required variables which have been introduced before, we arrive at
 \begin{eqnarray}
 {\partial {\hat \Delta F_{ns}}\over
	\partial \tau}(\nu,\tau)=\int_0^\nu{\Delta \hat F}_{ns}(w,\tau)
e^{-(\nu-w)}\left(\Delta P_{qq}^{LO,ns}(\nu-w)\right. \nonumber\\
\left. +{\alpha_s(\tau)\over 4\pi}\Delta
P_{qq}^{NLO,ns}(\nu-w)\right)d\,w\label{nonsingletinQsq_2}\;,\nonumber\\
\end{eqnarray}
which is in fact the extended version of Eq.(\ref{eq:ns}) for $\hat\Delta F_{ns}$.

Taking above equation to Laplace $s$-space, we obtain a linear differential equation in terms of $\tau$ variable for the transformed $\Delta f_{ns}(s,\tau)$. This equation has the simple solution as:
\begin{eqnarray}
 \Delta f_{NS}(s,\tau)=e^{\tau
	\Delta\Phi_{NS}(s)}\Delta f^{0}_{NS}(s),\nonumber\\ \qquad
\Delta\Phi_{NS}(s)\equiv \Delta\Phi_{NS}^{LO}(s)+{\tau_2\over
	\tau}\Delta\Phi_{NS}^{NLO}(s),\label{solutionandPhinsofs}
\end{eqnarray}
where
\begin{eqnarray}
\tau_2&\equiv& {1\over
	4\pi}\int_0^\tau\alpha_s(\tau')\,d\tau' ={1\over
	(4\pi)^2}\int_{Q^2_0}^{Q^2}\alpha_s^2(Q'^2)\, d \ln
Q'^2,\label{tau2ofQsq}\nonumber\\
\end{eqnarray}
and
\begin{eqnarray}
\Delta\Phi_{NS}^{LO}(s)\equiv{\cal L}\left [e^{-v}\Delta
P_{qq}^{LO,ns}(e^{-\nu});s\right ],\nonumber\\
\qquad\Delta\Phi_{NS}^{NLO}(s)\equiv{\cal L}\left [e^{-\nu}\Delta
P_{qq}^{NLO,NS}(e^{-\nu});s\right ].\label{PhiNs0and1}
\end{eqnarray}
It should be noted that at the LO approximation $\Delta\Phi_{NS}(s)=\Delta\Phi_f^{LO}(s)$ where $\Delta \Phi_f^{LO}(s)$ has been given explicitly in
Eq.(\ref{eq:qq}). The evaluation of $\Delta\Phi_{NS}^{NLO}(s)$ has been presented in Ref.\cite{AtashbarTehrani:2013qea}.
%----------------------------------
\begin{figure}[!htb]
		\vspace{-3.5cm}
	\includegraphics[clip,width=0.5\textwidth]{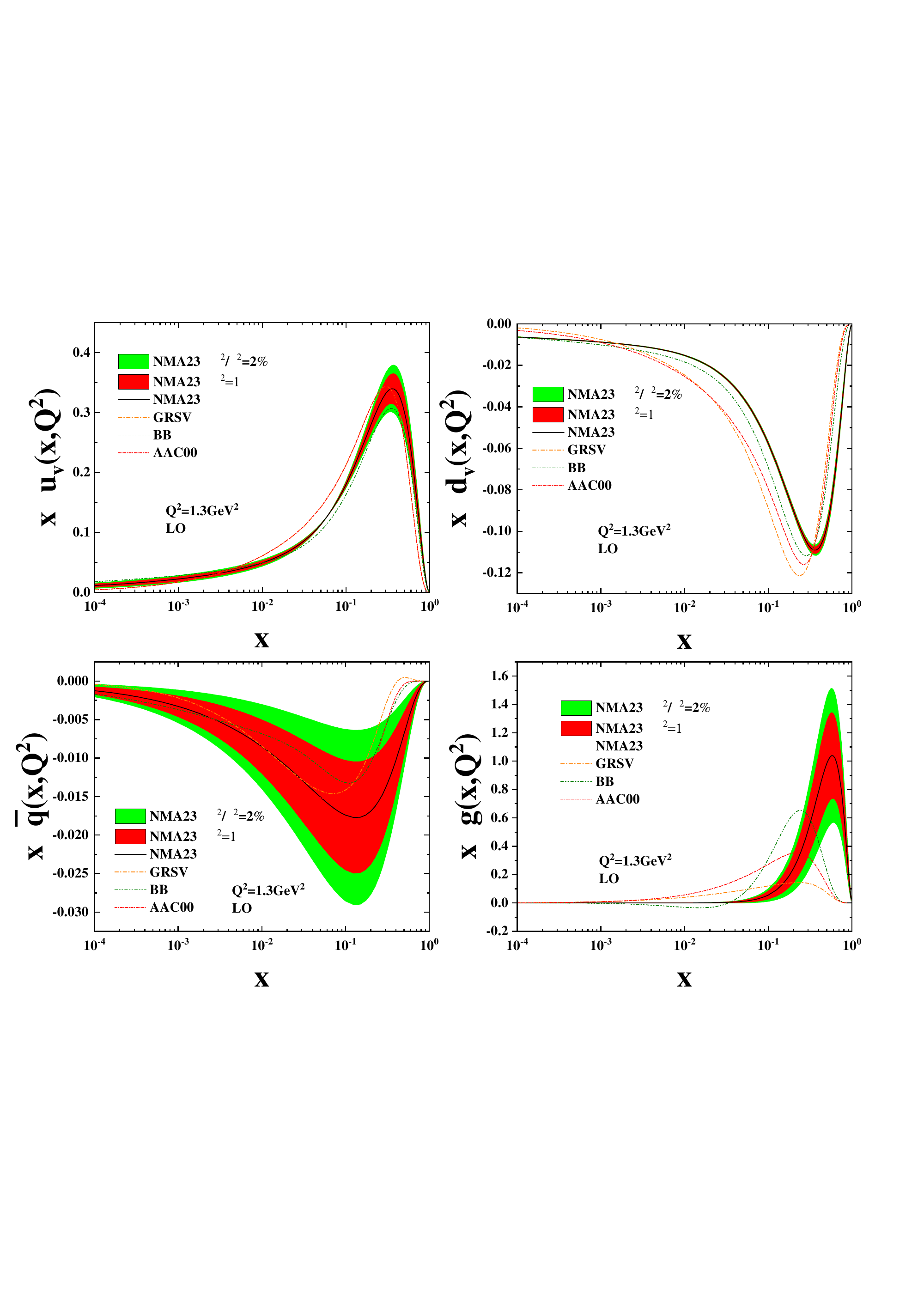}
	\vspace{-3.5cm}
	\begin{center}
		\caption{{\small Our NMA23  results for the polarized PDFs at Q$_0^2=$ 1.3 GeV$^2$
				with respect to $x$ in LO approximation, which is  plotted by a solid curve along with their $\Delta\chi^2/\chi^2 =2\%$
				uncertainty bands computed with the Hessian approach.
				We also present the result
				obtained in earlier global analyses of  BB (dashed)~\cite{Blumlein:2002qeu}, GRSV~(dashed-dotted)
				\cite{Gluck:2000dy}, AAC00 (dashed-dashed-dotted)~\cite{AsymmetryAnalysis:1999gsr} in LO approximation.  \label{fig:partonLOQ0}}}
	\end{center}
\end{figure}
%----------------------------------
%----------------------------------
\begin{figure}[!htb]
		\vspace{-3.5cm}
	\includegraphics[clip,width=0.5\textwidth]{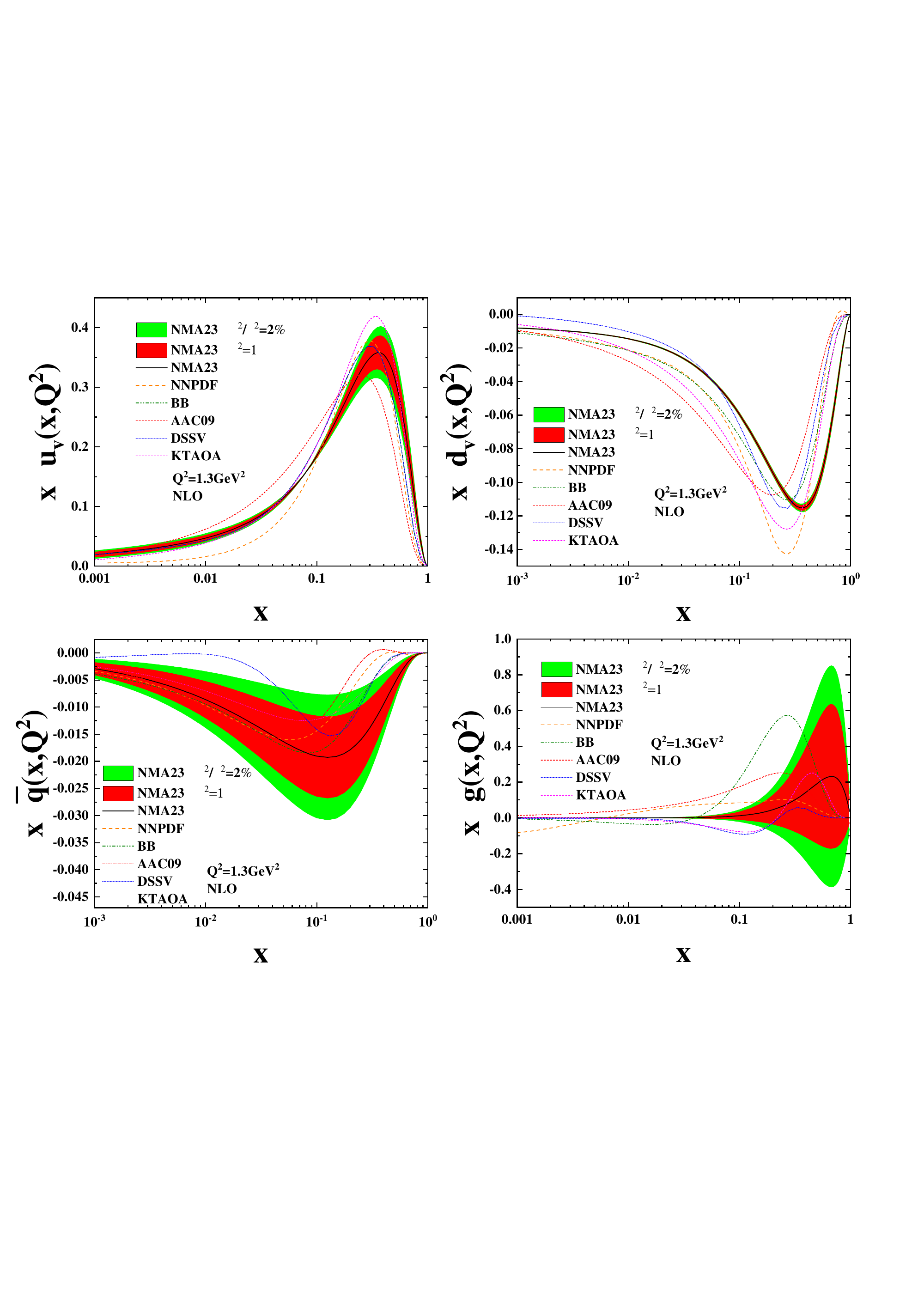}
	\vspace{-3.5cm}
	\begin{center}
		\caption{{\small Our NMA23  results for the spin-dependent PDFs at Q$_0^2=$ 1.3 GeV$^2$
				with respect to $x$ in NLO approximation which is plotted by a solid curve along with their $\Delta\chi^2/\chi^2 =2\%$
				uncertainty bands computed with the Hessian approach.
				We also display the result
				obtained in earlier global analyses of NNPDF(dashed-dotted-dotted)~\cite{Nocera:2014gqa}, KATAO (long dashed)~\cite{Khorramian:2010qa}, BB (dashed)~\cite{Blumlein:2010rn}, DSSV~(dashed-dotted)
				\cite{deFlorian:2014yva}, AAC09 (dashed-dashed-dotted)~\cite{Hirai:2008aj} in NLO approximation.  \label{fig:partonNLOQ0}}}
	\end{center}
\end{figure}
%----------------------------------
%----------------------------------
\begin{figure}[!htb]
	%	\vspace{-3.5cm}
	\includegraphics[clip,width=0.5\textwidth]{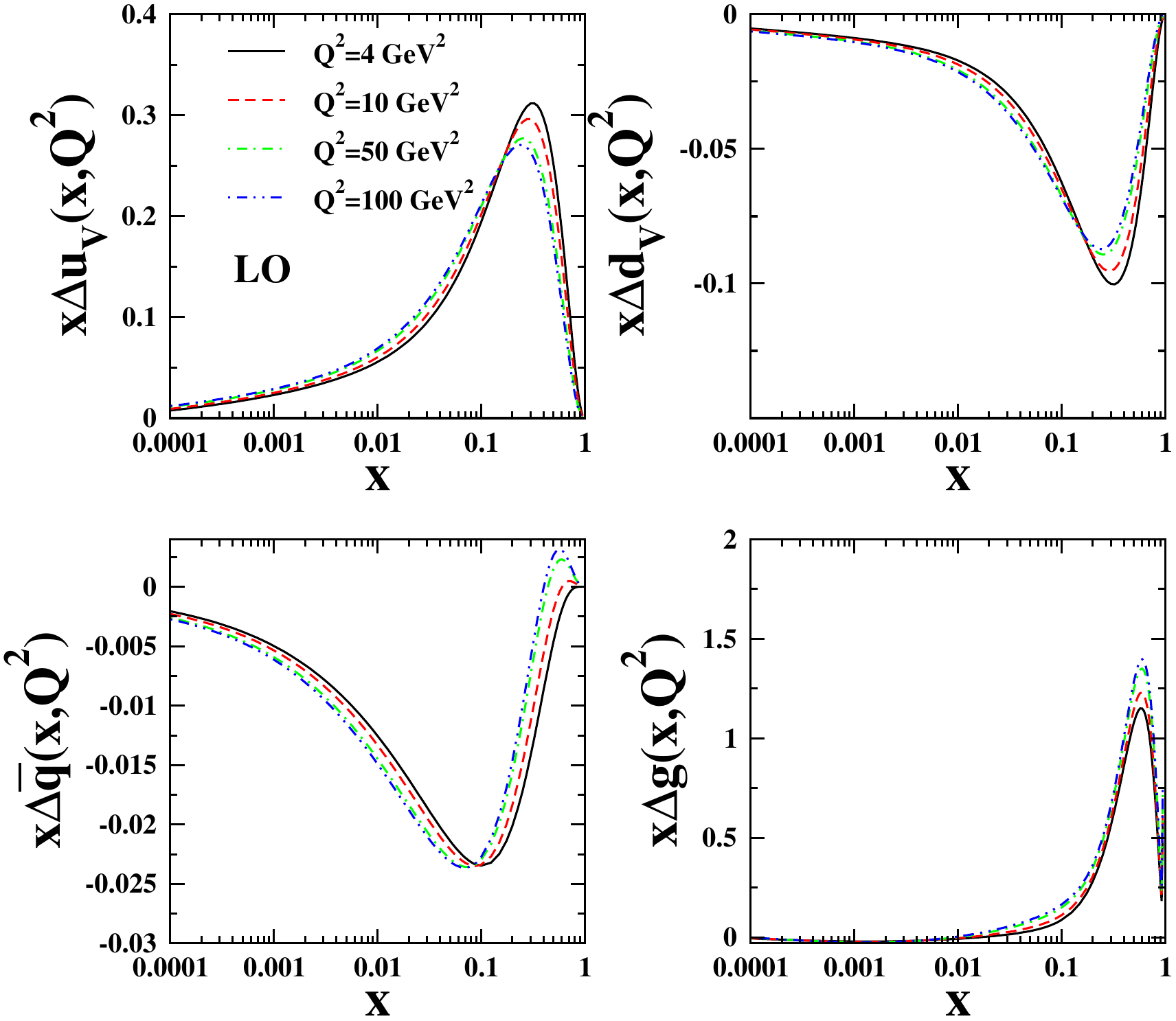}
	%\vspace{-6cm}
	\begin{center}
		\caption{{\small The evolved polarized quark densities as a function of $x$ in the LO
approximation.  \label{fig:partonLOev}}}
	\end{center}
\end{figure}
%----------------------------------
%----------------------------------
\begin{figure}[!htb]
	%	\vspace{-3.5cm}
	\includegraphics[clip,width=0.5\textwidth]{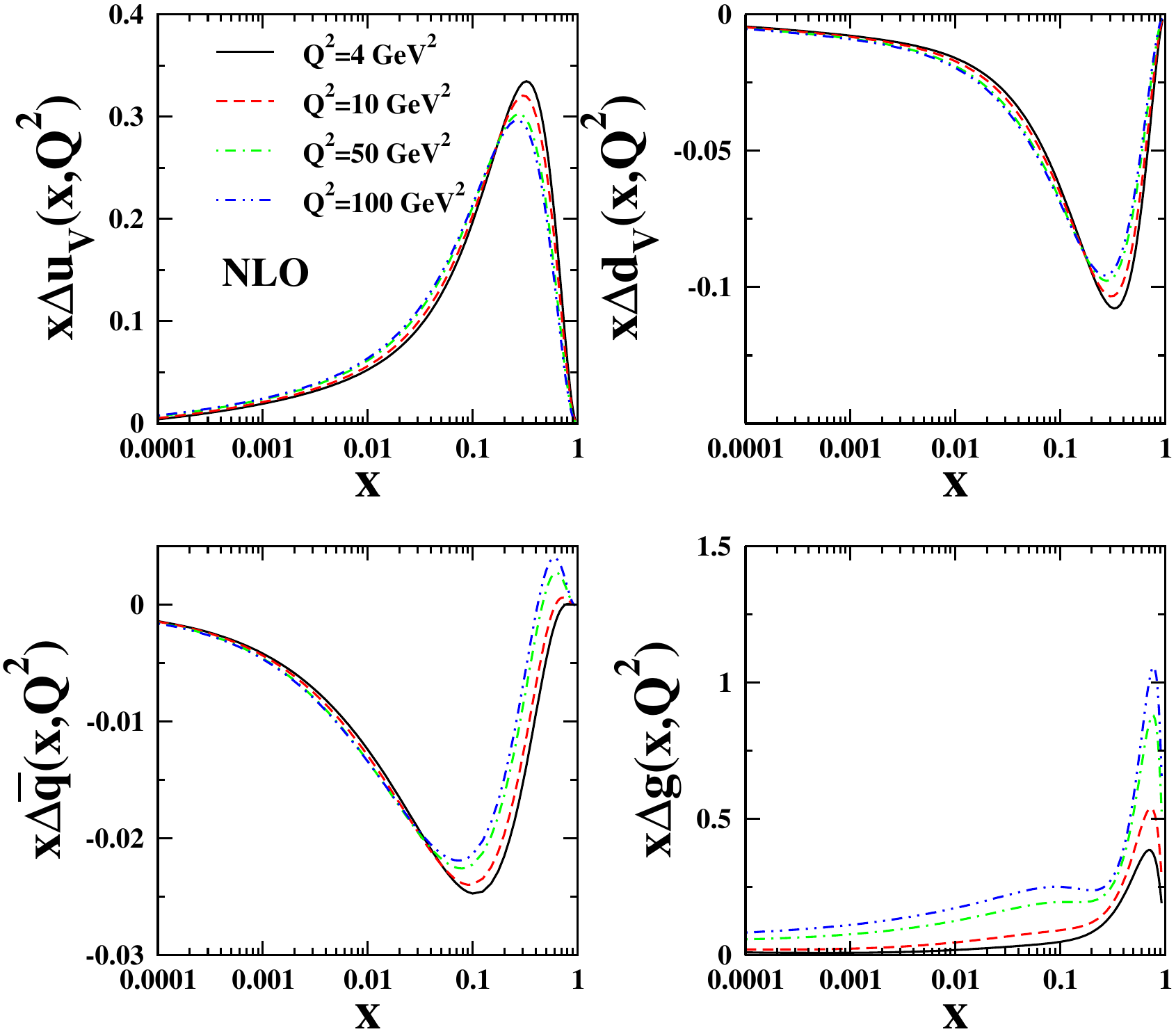}
	%\vspace{-6cm}
	\begin{center}
		\caption{{\small The evolved polarized quark densities as a function of $x$ in the NLO approximation. \label{fig:partonNLOev}}}
	\end{center}
\end{figure}
%----------------------------------

We can find any non-singlet solution, $\Delta F_{NS}(x,Q^2)$, by applying the non-singlet kernel $K_{ns}(v)\equiv {\cal L}^{-1}\left[e^{\tau \Delta\Phi_{NS}(s)};v \right]$, and using the Laplace convolution relation as:
\begin{eqnarray}
{ \Delta \hat
	F}_{NS}(v,\tau)=\int _0^v K_{ns}(v-w,\tau)\Delta \hat
F^{0}_{NS}(w)\,dw\label{Fnsofv}\;.
\end{eqnarray}
Taking the variable change $\nu\equiv\ln (1/x)$, the results are obtained in $(x,Q^2)$ space as before with this difference that the splitting functions in $s$-space include as well the NLO contributions in correspond to the following expressions:
\begin{eqnarray}
\Delta\Phi_S(s)\equiv \Delta\Phi_S^{LO}(s)+a_0\Delta\Phi_S^{NLO}(s),\nonumber\\
\Delta\Phi_g(s)\equiv \Delta\Phi_g^{LO}(s)+a_0\Delta\Phi_g^{NLO}(s),\label{Phis}\nonumber\\
\Delta\Theta_S(s)\equiv
\Delta\Theta_S^{LO}(s)+a_0\Delta\Theta_S^{NLO}(s),\nonumber\\
\Delta\Theta_g(s)\equiv
\Delta\Theta_g^{LO}(s)+a_0\Delta\Theta_g^{NLO}(s)\;.\label{Thetas}
\end{eqnarray}
The analytical expressions for $\Delta\Phi_{NS}^{NLO}(s)$ , $\Delta\Phi_S^{NLO}(s)$ , $\Delta\Phi_g^{NLO}(s)$, $\Delta\Theta_S^{NLO}(s)$ and $\Delta\Theta_g^{NLO}(s)$ in Laplace transform $s$-space have been presented in Ref.\cite{AtashbarTehrani:2013qea}.\\

We indicate the NLO expression for ${\alpha_s(\tau)\over 4\pi}$ by $a(\tau)$. We can numerically show that an excellent approximation to $a(\tau)\equiv {\alpha_s(\tau)\over 4\pi} $, with a precision of a few parts in $10^4$, is given by the expression
\begin{eqnarray}
 a(\tau)&\approx
a_0+ a_1e^{-b_1\tau}, \label{aoftau}
\end{eqnarray}
where the unknown parameters $a_1, b_1$ and $a_0$ are found by a least squared fit to $a(\tau)$. It should be pointed that this approximation is inspired by the fact that at the LO approximation the expression for $\alpha_{s,LO}(\tau)$ is exactly given by $\alpha_{s,LO}(Q_0^2)e^{-b \tau}$.\\

Now for the singlet sector and gluon part of  Eq.(\ref{LO-f}) at the NLO approximation, the results of DGLAP evolution  equations in Laplace s-space could be obtained and are given by:
\begin{eqnarray}
\Delta f(s,\tau)&=&k_{ff}(a_1,b_1,s,\tau)\Delta f^0_S(s)+k_{fg}(a_1,b_1,s,\tau)\Delta g^0(s),\nonumber\\
\Delta g(s,\tau)&=&k_{gg}(a_1,b_1,s,\tau)\Delta g^0(s)+k_{gf}(a_1,b_1,s,\tau)\Delta f^o_S(s)\label{fandg}\;.\nonumber\\
\end{eqnarray}
Here the functions $k_{ij}(a_1,b_1,s,\tau)$ are expressed as a power series in terms of the NLO expansion parameter $a_1$  whose coefficients are analytic functions with respect to $s$ and $\tau$ parameters. These expressions which are used  to extract polarized parton distribution functions at NLO approximation, can be found in Ref.\cite{AtashbarTehrani:2013qea}.
Finally, recalling that $\nu\equiv \ln(1/x)$, we can convert the above solutions to the $(x,Q^2)$ space. Therefore we should write the NLO decoupled solutions, $\Delta F_s(x,Q^2)$ and $\Delta G(x,Q^2)$, with the knowledge requirement of $\Delta F(x)$ and $\Delta G(x)$ at initial scale $Q_0^2$. We also use these analytical solutions for polarized parton distributions in the
next sections to extract polarized structure functions of proton, neutron and deuteron.
\begin{table*}[htb]
	\caption{Summary of published polarized DIS experimental data points with measured $x$ and $Q^2$ ranges and the number of data points.} \label{tab:DISdata}
	%	\begin{ruledtabular}
	\begin{tabular}{l c c c c c c c}
		\hline
		\textbf{Experiment} & \textbf{Ref.} & \textbf{[$x_{\rm min}, x_{\rm max}$]}  & \textbf{Q$^2$  {(}GeV$^2${)}}  & \textbf{data poi.} & $\chi^2_{LO}$ & $\chi^2_{NLO}$ & ${\cal N}_i$     \tabularnewline
		\hline\hline
		\textbf {SLAC/E143(p)}   & \cite{Abe:1998wq}   & [0.031--0.749]   & 1.27--9.52 & 28& 25.9418 & 25.5862 &0.997894385554395  \\
		\textbf{HERMES(p)} & \cite{HERM98}  & [0.028--0.66]    & 1.01--7.36 & 39& 57.6213 &  54.4579 &1.00121847577922 \\
		\textbf{SMC(p)}    & \cite{Adeva:1998vv}    & [0.005--0.480]   & 1.30--58.0 & 12& 5.7338 &  7.807 &0.999682542198503 \\
		\textbf{EMC(p)}    & \cite{Ashman:1987hv}     & [0.015--0.466]   & 3.50--29.5 & 10& 5.5308 & 5.2507 &0.998746870048074 \\
		\textbf{SLAC/E155}      & \cite{E155p}    & [0.015--0.750]   & 1.22--34.72 & 24& 30.8939 & 25.8174 &1.00866491330750 \\
		\textbf{HERMES06(p)} & \cite{HERMpd} & [0.026--0.731]   & 1.12--14.29 & 51& 25.1976 & 30.7736 &0.978727363512146 \\
		\textbf{COMPASS10(p)} & \cite{COMP1} & [0.005--0.568]   & 1.10--62.10 & 15& 21.1180 & 21.7614 &0.986930436542322\\
		\textbf{COMPASS16(p)} & \cite{Adolph:2015saz} & [0.0035--0.575]   & 1.03--96.1 & 54& 38.1731 & 36.7525 &0.998144618082626 \\
		\textbf {SLAC/E143(p)}   & \cite{Abe:1998wq}   & [0.031--0.749]   & 2-3-5 & 84& 108.5918 &102.0200 &0.998508244589941  \\
		\textbf{HERMES(p)} & \cite{HERM98}  & [0.023--0.66]    &      2.5    & 20& 34.3329 & 32.2223 &1.00223612389713  \\
		\textbf{SMC(p)}    & \cite{Adeva:1998vv}    & [0.003--0.4]   & 10 & 12& 10.2414 & 10.2829 &1.00050424220605 \\
		\textbf{Jlab06(p)}&\cite{Dharmawardane:2006zd} & [0.3771--0.9086]     &3.48--4.96 & 70& 101.8636 &103.0134 &0.999973926530214 \\
		\textbf{Jlab17(p)}&\cite{Fersch:2017qrq} & [0.37696--0.94585]     &3.01503--5.75676     & 82&  178.2875 & 186.4491 &1.00223612389713 \\
		\multicolumn{1}{c}{$\bf{g_1^p}$}         &  &  &   &  \textbf{501} && \\
		\textbf{SLAC/E143(d)}  &\cite{Abe:1998wq}    & [0.031--0.749]   & 1.27--9.52    & 28& 37.8573 & 38.4526&1.00123946481403\\
		\textbf{SLAC/E155(d)}  &\cite{E155d}     & [0.015--0.750]   & 1.22--34.79   & 24& 19.6385 & 18.3428 &1.00093216947938 \\
		\textbf{SMC(d)}   &\cite{Adeva:1998vv}     & [0.005--0.479]   & 1.30--54.80   & 12& 19.1969 &18.8501 &1.00004243740921 \\
		\textbf{HERMES06(d)} & \cite{HERMpd}& [0.026--0.731]   & 1.12--14.29   & 51& 48.9197 & 48.3606 &1.00287123838967  \\
		\textbf{COMPASS05(d)}& \cite{COMP2005}& [0.0051--0.4740] & 1.18--47.5   & 11& 8.0128 & 8.6001 &1.00103525695298 \\
		\textbf{COMPASS06(d)}& \cite{COMP2006}& [0.0046--0.566] & 1.10--55.3    & 15& 5.4623 & 7.6560 &1.00014044224998 \\
		\textbf{COMPASS17(d)} & \cite{Adolph:2016myg} & [0.0045--0.569]   & 1.03--74.1 & 43& 33.1002 &31.4721 &1.00401646687751   \\
		\textbf{SLAC/E143(d)}  &\cite{Abe:1998wq}    & [0.031--0.749]   & 2--3--5    & 84& 125.8333 &125.7379 & 0.999955538651217\\
		\multicolumn{1}{c}{ $\bf{g_1^d}$}        &  & & & \textbf{268}  & &   \\
		\textbf{SLAC/E142(n)}   &\cite{E142n}    & [0.035--0.466]   & 1.10--5.50    & 8& 7.9561 & 7.7586 &0.998697021286838 \\
		\textbf{HERMES(n)} &\cite{HERM98}   & [0.033--0.464]   & 1.22--5.25    & 9&  2.3999 & 2.5486 & 0.999948762872958 \\
		\textbf{E154(n)}   &\cite{E154n}    & [0.017--0.564]   & 1.20--15.00   & 17& 24.3141 & 21.4152 &0.999115473491324 \\
		\textbf{HERMES06(n)} &\cite{Ackerstaff:1997ws}  &  [0.026--0.731]  & 1.12--14.29   & 51& 18.2773 & 17.3811 &0.998906625804611  \\
		\textbf{Jlab03(n)}&\cite{JLABn2003} & [0.14--0.22]     & 1.09--1.46    & 4&5.6144e-2 &5.7129e-2 & 0.999554786214114 \\
		\textbf{Jlab04(n)}&\cite{JLABn2004} & [0.33--0.60]      & 2.71--4.8     & 3&14.5393 &8.8133 &0.994389736514520\\
		\textbf{Jlab05(n)}&\cite{JLABn2005} & [0.19--0.20]     &1.13--1.34     & 2&7.7029 & 6.7595& 0.999939246942750  \\
		\multicolumn{1}{c}{$\bf{g_1^n}$}     &  &     & & \textbf{94} & &  \\
		\textbf{E143(p)}    & \cite{Abe:1998wq}   & [0.038--0.595]   & 1.49--8.85    & 12&11.1698& 10.9170&1.00335621969896 \\
		\textbf{E155(p)}   &\cite{E155pdg2}  &[0.038--0.780]    & 1.1--8.4      & 8 &12.8132&15.6826 &1.04042312148245 \\
		\textbf{Hermes12(p)}&\cite{hermes2012g2} &[0.039--0.678]&1.09--10.35   & 20&25.1271&21.5690 &1.00274614220085 \\
		\textbf{SMC(p)}      &\cite{SMCpg2} & [0.010--0.378]    & 1.36--17.07   & 6 &1.8259& 1.7117 &1.00001538561406  \\
		\multicolumn{1}{c}{$\boldsymbol{g_2^p}$}  &  &  & & \textbf{46} &    \\
		\textbf{E143(d)}     &\cite{Abe:1998wq} & [0.038--0.595]   & 1.49--8.86    & 12 &9.6009&9.6132 &1.00047527921467 \\
		\textbf{E155(d)}    &\cite{E155pdg2}& [0.038--0.780]    & 1.1--8.2      & 8 &12.2433&12.275 &1.01388492611179  \\
		\multicolumn{1}{c}{$\boldsymbol{g_2^d}$}  &  &  & & \textbf{20} &   \\
		\textbf{E143(n)}    &\cite{Abe:1998wq}  & [0.038--0.595]   & 1.49--8.86    & 12&8.9660& 9.0283  &1.00004452228917 \\
		\textbf{E155(n)}    &\cite{E155pdg2}&[0.038--0.780]    &1.1--8.8       & 8 &14.1622& 13.6977 &1.03135886789238 \\
		\textbf{E142(n)}    &\cite{E142n}   &[0.036--0.466]    &1.1--5.5       & 8 &16.48220&3.8883 &1.00000431789744 \\
		\textbf{Jlab03(n)}  &\cite{JLABn2003}&[0.14--0.22]     & 1.09--1.46    & 4 &17.6171&13.8173 & 1.03226263480608  \\
		\textbf{Jlab04(n)}  &\cite{JLABn2004}&[0.33--0.60]     & 2.71--4.83    & 3 &4.2782&4.4079 &0.900030714490705 \\
		\textbf{Jlab05(n)}  &\cite{JLABn2005}&[0.19--0.20]     & 1.13--1.34    & 2 &10.1400&8.0260 &0.981366577296903 \\
		\multicolumn{1}{c}{$\boldsymbol{g_2^n}$}  &  &  & &\textbf{37} &   \\  \hline
		\hline\\
		\multicolumn{1}{c}{\textbf{ Total}}&\multicolumn{4}{c}{~~~~~~~~~~~~~~~~~~~~~~~~~~~~~~~~~~~~~~~~~~~~~~~~~~~~~~~\textbf{966}}&\multicolumn{1}{c}{\textbf{ 1171.6502}}&\multicolumn{1}{c}{\textbf{1128.9857}}
		\\
		\hline\\
	\end{tabular}
	%	\end{ruledtabular}
\end{table*}
\section{The Jacobi Polynomial Method \label{sec:4}}

We perform a fit in the LO approximation for the polarized parton distributions using Jacobi polynomials \cite{Kataev:1997nc,Kataev:1998ce,Kataev:1999bp,Kataev:2001kk} to reconstruct the $x$ dependent quantities from their Laplace moments. The application of Jacobi polynomials has a number of advantages; especially, it will provide us an opportunity to factorize out the $x$and $Q^{2}$ dependence which help us to have an efficient parametrization and to do the evolution of the structure functions.

For example, we can expand the spin structure function $xg_{1}(x,Q^{2})$, as \cite{Kataev:1999bp}:
\begin{equation}
xg_{1}(x,Q^{2})=x^{\beta}(1-x)^{\alpha}\ \sum_{n=0}^{N_{max}}a_{n}(Q^{2})\ \Theta_{n}^{\alpha,\beta}(x)\,\label{eq:xg1}\end{equation}

in which $\Theta_{n}^{\alpha,\beta}(x)$ are Jacobi polynomials of order $n$, and $N_{max}$ is the maximum order of the expansion. In this case, the $Q^{2}$-dependence of the polarized structure function is contained in the Jacobi moments, $a_{n}(Q^{2})$. On the other hand, we can factor out the essential part of its $x$-dependence into a weight function using the Jacobi polynomials \cite{parisi}.

For computational purpose, the $x$-dependence of the Jacobi polynomials is given by the following expansion \cite{Kataev:2001kk}:
\begin{equation}
\Theta_{n}^{\alpha,\beta}(x)=\sum_{j=0}^{n}c_{j}^{(n)}(\alpha,\beta)\ x^{j},\label{eq:jac}
\end{equation}
where the $c_{j}^{(n)}(\alpha,\beta)$'s are combinations of $\Gamma$-functions. The Jacobi polynomials satisfy an orthogonality condition with weight function $x^{\beta}(1-x)^{\alpha}$ such that:

\begin{equation}
\int_{0}^{1}dx\; x^{\beta}(1-x)^{\alpha}\Theta_{k}^{\alpha,\beta}(x)\Theta_{l}^{\alpha,\beta}(x)=\Delta_{k,l}\ .\label{eq:ortho}
\end{equation}

Hence, the polarized structure function $xg_{1}(x,Q^{2})$ could be reconstructed from Eq.~(\ref{eq:xg1}), by giving the Jacobi moments $a_{n}(Q^{2})$ \cite{Shahri:2016uzl,Khanpour:2017cha,Khanpour:2017fey,Khorramian:2010qa,Khorramian:2009xz,MoosaviNejad:2016ebo,Khanpour:2016uxh,Nematollahi:2021ynm,Mirjalili:2022cal,AtashbarTehrani:2013qea}.

We can obtain the Jacobi moments $a_{n}(Q^{2})$, taking the orthogonality condition on Eq.~(\ref{eq:xg1}) which finally lead us to:
 \begin{eqnarray}
a_{n}(Q^{2})
& = & \sum_{j=0}^{n}c_{j}^{(n)}(\alpha,\beta)\ {\cal L}[xg_{1},s=j+1]~\ .\label{eq:aMom}
\end{eqnarray}

In deriving Eq.~(\ref{eq:aMom}), we utilize the Laplace transform of $xg_{1}(x,Q^{2})$ as it follows:

\begin{eqnarray}
{\cal {L}}[xg_{1},s] & \equiv & \int_{0}^{\infty}dv\ e^{-sv}\ xg_{1}(x,Q^{2})\ .\label{eq:Mellin}
\end{eqnarray}
We can now relate the polarized structure function, $xg_{1}(x,Q^{2})$,
with its moments in Laplace s-space as following \cite{Shahri:2016uzl,Khanpour:2017cha,Khanpour:2017fey,Khorramian:2010qa,Khorramian:2009xz,MoosaviNejad:2016ebo,Khanpour:2016uxh,Nematollahi:2021ynm,Mirjalili:2022cal,AtashbarTehrani:2013qea}

\begin{eqnarray}
xg_{1}(x,Q^{2}) & = & x^{\beta}(1-x)^{\alpha}\sum_{n=0}^{N_{max}}\Theta_{n}^{\alpha,\beta}(x)\nonumber \\
& \times & \sum_{j=0}^{n}c_{j}^{(n)}{(\alpha,\beta)}\ {\cal {L}}[xg_{1} ,s=j+1]\ .\label{eg1Jacob}
\end{eqnarray}

By regarding Eq.~(\ref{eg1Jacob}) for $xg_{1}(x,Q^{2})$, we choose the set $\{N_{max},\alpha,\beta\}$ to reach optimal convergence of this series throughout the kinematic region constrained by the data.
In practice, we find the following numerical values for above parameters: $N_{max}=9$, $\alpha=3.0$, and $\beta=0.5$ to be sufficient. We should note that in Mellin space when we intend to calculate the first moment for quark distribution or structure function, $g_1$,  we choose $n=1$ but in Laplace s-space to get the first moment, we should consider $s=0$ instead of.
\section{QCD Analysis \& Parametrization of PPDFs\label{sec:5}}
The required analysis of PPDFs within the QCD content, are including the following parts.
\subsection{Parametrization}
We consider a proton consisted of massless partons which carry momentum fraction $x$ with helicity distributions $q_{\pm}(x,Q^{2})$ at characteristic scale $Q^2$. The difference $\Delta q(x,Q^{2})=q_{+}(x,Q^{2})-q_{-}(x,Q^{2})$ measures how much the parton of flavor $q$ remembers its parent's proton polarization. On the other words we can say that it represents the probability of finding a polarized parton with fraction $x$ of parent hadron momentum and spin align/anti-align to hadron's spin. It measures the net helicity of partons in a longitudinally polarized hadron.

In parametrization process, we consider the following form for the polarized PDFs at initial scale $Q_{0}^{2}=1.3$ GeV$^{2}$: \begin{equation}
x\:\Delta q(x,Q_{0}^{2})={\cal N}_{q}\eta_{q}x^{a_{q}}(1-x)^{b_{q}}(1+c_{q}x)\ ,\label{eq:parm}
\end{equation}
in which the polarized PDFs are determined by parameters $\{\eta_{q},a_{q},b_{q},c_{q}\}$, and the generic label $q=\{u_{v},d_{v},\bar{q},g\}$ indicates the partonic flavors up-valence, down-valence, sea, and gluon, respectively. ${\cal N}_{q}$ is the normalization constant given by:
\begin{equation}
\frac{1}{{\cal N}_{q}}=\left(1+c_{q}\frac{a_{q}}{a_{q}+b_{q}+1}\right)\, B\left(a_{q},b_{q}+1\right)\ ,\label{eq:norm}
\end{equation}
and chosen such that $\eta_{q}$ in Eq.(\ref{eq:parm}) is the first moments of $\Delta q(x,Q_{0}^{2})$ where $B(a,b)$ is the Euler beta function.

The total up and down quark distributions are a sum of the valence plus sea distributions:
$\Delta u=\Delta u_{v}+\Delta\bar{q}$ and $\Delta d=\Delta d_{v}+\Delta\bar{q}$.
We consider an $SU(3)$ flavor symmetry as $\Delta\overline{q}\equiv\Delta\overline{u}=\Delta\overline{d}=\Delta s=\Delta\overline{s}$. Nevertheless we could allow for an $SU(3)$ symmetry violation term by introducing $\kappa$ such that $\Delta s=\Delta\overline{s}=\kappa\Delta\bar{q}$.
Since the strange quark distribution is poorly constrained, the results would be insensitive to the particular choice of $\kappa$.

From Eq.~(\ref{eq:parm}), it is obvious that each of four polarized parton densities $q=\{u_{v},d_{v},\bar{q},g\}$ contain four parameters $\{\eta_{q},a_{q},b_{q},c_{q}\}$ which gives a total of 16 parameters that should be determined. We illustrate that some of these parameters can be eliminated while maintaining sufficient flexibility to obtain a good fit.
%----------------------------------
\begin{figure}[!htb]
	%	\vspace{-3.5cm}
	\includegraphics[clip,width=0.5\textwidth]{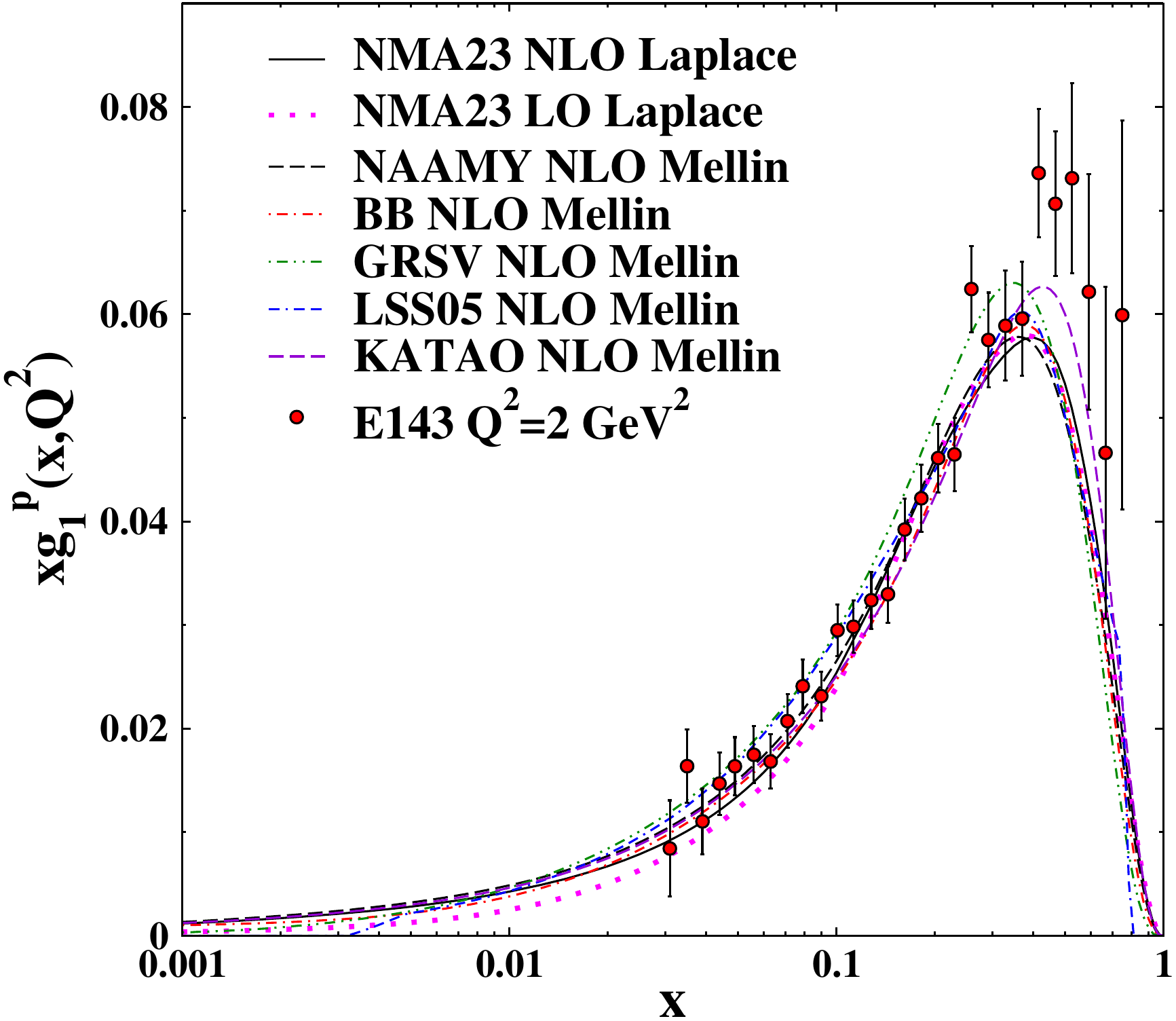}
	%\vspace{-6cm}
	\begin{center}
		\caption{{\small  The polarized proton structure functions with respect to $x$ at $Q^2=2\;GeV^2$. The results of Jacobi expansion technique in NLO (solid curve) and LO (dotted) approximations are compared with parametrization models such as NAAMY (long-dashed)~\cite{Nematollahi:2021ynm}, BB (dashed-dotted) \cite{Blumlein:2010rn}, GRSV (dashed-dotted-dotted) \cite{Gluck:2000dy}, LSS05 (dashed-dashed-dotted) \cite{Leader:2005ci} and KATAO (dashed)~\cite{Khorramian:2010qa}.  \label{fig:xg1pE143}}}
	\end{center}
\end{figure}
%----------------------------------
%----------------------------------
\begin{figure}[!htb]
	%	\vspace{-3.5cm}
	\includegraphics[clip,width=0.5\textwidth]{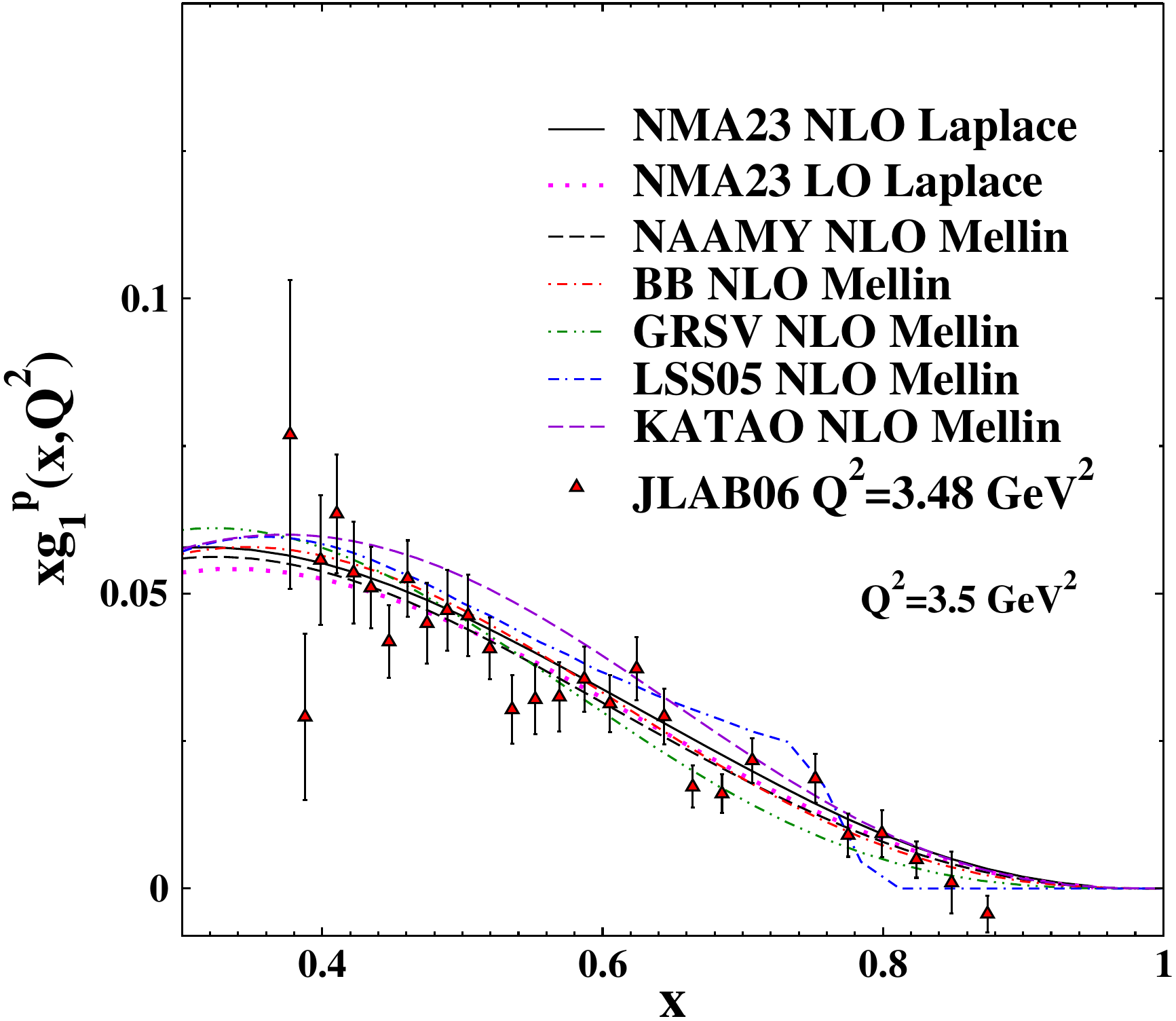}
	%\vspace{-6cm}
	\begin{center}
		\caption{{\small As in Fig.\ref{fig:xg1pE143} but at $Q^2=3.5\;GeV^2$ and in a shorter range of $x$, corresponding to the available data. \label{fig:xg1pjlab06}}}
	\end{center}
\end{figure}
%----------------------------------

%----------------------------------
\begin{figure}[!htb]
	%	\vspace{-3.5cm}
	\includegraphics[clip,width=0.5\textwidth]{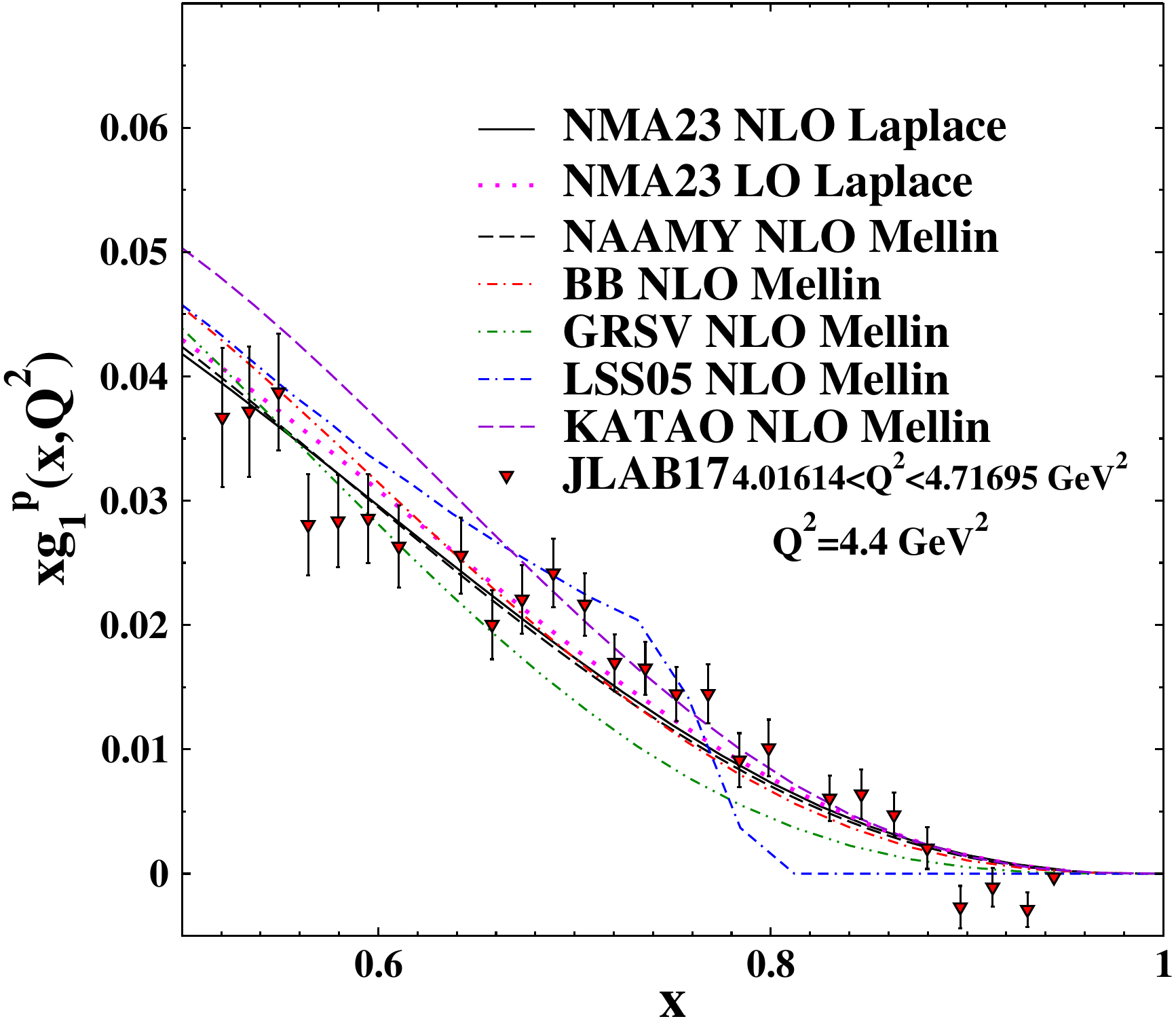}
	%\vspace{-6cm}
	\begin{center}
		\caption{{\small As in Fig.\ref{fig:xg1pE143} but at $Q^2=4.4\;GeV^2$ and in a shorter range of $x$, corresponding to the available data.  \label{fig:xg1pjlab17}}}
	\end{center}
\end{figure}
%----------------------------------

%----------------------------------
\begin{figure}[!htb]
	%	\vspace{-3.5cm}
	\includegraphics[clip,width=0.5\textwidth]{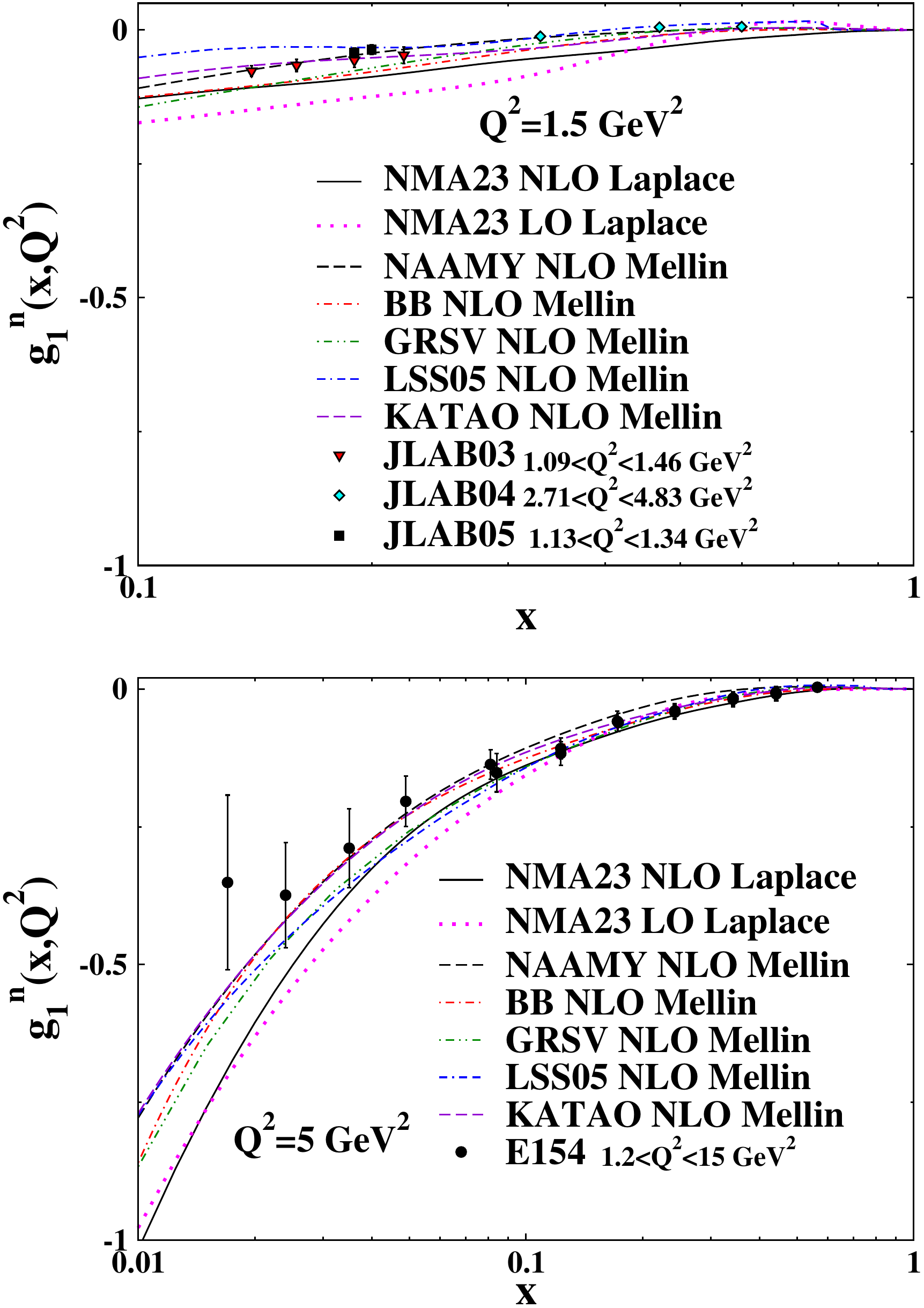}
	%\vspace{-6cm}
	\begin{center}
		\caption{{\small The spin dependent neutron structure functions with respect to $x$ at $Q^2$ $=1.5\;GeV^2$ and $5\;GeV^2$. The results of Jacobi expansion technique in NLO (solid curve) and LO (dotted) approximations are compared with parametrization models like NAAMY (long-dashed)~\cite{Nematollahi:2021ynm}, BB (dashed-dotted) \cite{Blumlein:2010rn}, GRSV (dashed-dotted-dotted) \cite{Gluck:2000dy}, LSS05 (dashed-dashed-dotted) \cite{Leader:2005ci} and KATAO (dashed)~\cite{Khorramian:2010qa}.   \label{fig:g1njlab}}}
	\end{center}
\end{figure}
%----------------------------------

%----------------------------------
\begin{figure}[!htb]
	%	\vspace{-3.5cm}
	\includegraphics[clip,width=0.5\textwidth]{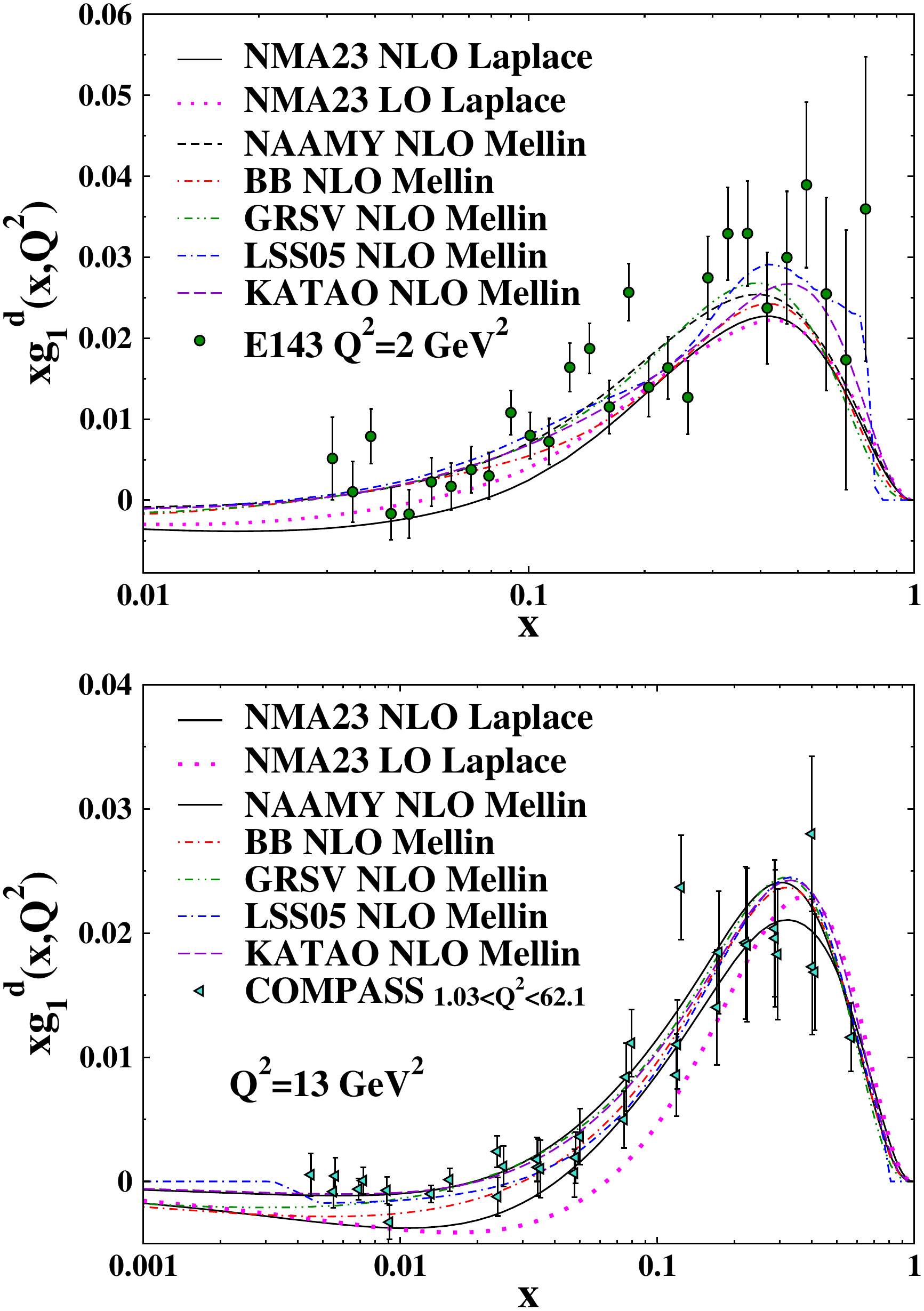}
	%\vspace{-6cm}
	\begin{center}
		\caption{{\small The spin dependent structure functions of deuteron as a function of $x$ at $Q^2$ $=2\;GeV^2$ and $13\;GeV^2$. The results of Jacobi expansion technique in NLO (solid curve) and LO (dotted) approximations are compared with  parametrization models such as NAAMY (long-dashed)~\cite{Nematollahi:2021ynm}, BB (dashed-dotted) \cite{Blumlein:2010rn}, GRSV (dashed-dotted-dotted) \cite{Gluck:2000dy}, LSS05 (dashed-dashed-dotted) \cite{Leader:2005ci} and KATAO (dashed)~\cite{Khorramian:2010qa}.   \label{fig:xg1dcompass}}}
	\end{center}
\end{figure}
\subsection{First Moments of $\Delta u_{v}$ and $\Delta d_{v}$}

The parameters $\eta_{u_{v}}$ and $\eta_{d_{v}}$ are the first moments of the polarized valence up and down quark densities, denoted by $\Delta u_{v}$ and $\Delta d_{v}$.  These densities can be related to $F$ and $D$ quantities as the weak matrix elements which are measured in neutron and hyperon $\beta$--decays. Hence one can write \cite{PDG}:
\begin{eqnarray}
a_{3} & = & \int_{0}^{1}dx\:\Delta q_{3}=\eta_{u_{v}}-\eta_{d_{v}}=F+D\ ,\\
a_{8} & = & \int_{0}^{1}dx\:\Delta q_{8}=\eta_{u_{v}}+\eta_{d_{v}}=3F-D\ ,
\end{eqnarray}
where $a_{3}$ and $a_{8}$ denote the non-singlet combinations of the first moments of the polarized quark densities corresponding to
\begin{eqnarray}
q_{3} & = & (\Delta u+\Delta\overline{u})-(\Delta d+\Delta\overline{d})\ ,\\
q_{8} & = & (\Delta u+\Delta\overline{u})+(\Delta d+\Delta\overline{d})-2(\Delta s+\Delta\overline{s})\ .
\end{eqnarray}

\noindent A reanalysis of $F$ and $D$ with updated $\beta$-decay constants leads to the following results: $F=0.464\pm0.008$ and $D=0.806\pm0.008$  \cite{PDG}. With these values we obtain:

\begin{eqnarray}
\eta_{u_{v}} & = & +0.928\pm0.014\ ,\label{eq:etauv}\\
\eta_{d_{v}} & = & -0.342\pm0.018\ .\label{eq:etadv}
\end{eqnarray}

Utilizing the above numerical values of $\eta_{u_{v}}$ and $\eta_{d_{v}}$ will end to reduce two parameters during the fitting processes.
\subsection{Polarized DGLAP evolution}
The polarized DGLAP evolution equations can be solved in the Laplace space using the Jacobi polynomial approach. The Laplace transformation of the parton densities $\Delta q$ are defined analogous to that of Eq.~(\ref{eq:Mellin}) as:
\begin{eqnarray}
&{\cal{L}}&[\Delta q(x=e^{-v},Q_{0}^{2}),s]  \equiv \nonumber\\
\Delta q(s,Q_{0}^{2})&=&\int_{0}^{\infty}e^{-sv}\:\Delta q (x=e^{-v},Q_{0}^{2})\: dv\nonumber \\
& = & {\cal {N}}_{q}\eta_{q}\left(1+c_{q}\:\frac{s+a_{q}}{s+a_{q}+b_{q}+1}\right)\ \nonumber \\
& \times & B(s+a_{q},b_{q}+1)\ ,
\end{eqnarray}
where $q=\{u_{v},d_{v},\overline{q},g\}$, and $B$ denotes the Euler beta function.

The twist-2 contributions to the spin dependent structure function $g_{1}(s,Q^{2})$ can be expressed in terms of the polarized parton densities, in the Laplace space as:
\begin{eqnarray}
{\cal L}[g_{1}^{p},s] & =&  \frac{1}{2}\sum\limits _{q}e_{q}^{2}
\times \left[\left(1+\frac{\tau}{4\pi}\Delta C_q(s)\right) [\Delta q(s,Q^{2})+\right.\nonumber\\
&&\left.\Delta\bar{q}(s,Q^{2})]+\frac{2}{3}\frac{\tau}{4\pi}\Delta C_g(s) \Delta g(s,Q^{2})\right]\;,\label{g1ps}
\end{eqnarray}
\begin{eqnarray}
{\cal L}[g_{1}^{n},s]  &=& {\cal L}[g_{1}^{p},s]- \frac{1}{6} (
{\cal L}[u_{v},s] -\nonumber\\&&{\cal L}[d_{v},s])\times \left(1+\frac{\tau}{4\pi}\Delta C_q(s)\right)\;.\label{g1ns}
\end{eqnarray}
Here, the summation is over $ u,d,s $ quark flavors. $\delta q,\delta\bar{q}$ and $\delta g$ are the polarized quark, anti-quark, and gluon distributions, respectively. The $\Delta C_g$ and $\Delta C_q$ denote the spin dependent Wilson coefficients in Laplace transform $s$ space respectively which are written as:
\begin{eqnarray}
\Delta C_g&=&\frac{1}{2} \left(\frac{2}{s+1}-\frac{2}{s+2}+\frac{\psi ^{(0)}(s+1)+\gamma_E }{s+1}-\right.\nonumber\\
&&\left.\frac{2 \left(\psi ^{(0)}(s+2)+\gamma_E
	\right)}{s+2}\right)
\end{eqnarray}
\begin{eqnarray}
\Delta C_q&=&\frac{8}{3 (s+1)}+\frac{4}{3 (s+2)}+\frac{4 \left(\psi ^{(0)}(s+2)+\gamma_E \right)}{3 (s+1)}+\nonumber\\
&&\frac{4 \left(\psi
	^{(0)}(s+3)+\gamma_E \right)}{3 (s+2)}+\frac{4 \psi ^{(1)}(s+1)}{3}+\frac{4 \psi ^{(1)}(s+3)}{3}\nonumber\\&&
-\frac{4}{3}\left(\frac{9}{2}+\frac{\pi ^2}{3}\right)
\end{eqnarray}
Employing the inverse Laplace transform on Eqs.(\ref{g1ps},\ref{g1ns}), the $g_{1}^{p}$ and $g_{1}^{n}$ can be obtained in Bjorken $x$-space. Considering the parametrization for PPDFs there are finally 9 unknown parameters which should be determined during the fitting processes, taking the available experimental data for polarized structure functions.\\

In addition to proton and neutron structure functions we can also do the required computations for the deuteron which is in fact a nucleus consists of one proton and one neutron. The deuteron structure function is given by $xg_{i}^{d}=\frac{xg_{i}^{p}+xg_{i}^{n}}{2}\times (1 - 1.5 \omega_D)\; ; i=1,2$ . Here $ \omega  _D=0.05\pm0.01$ denotes the probability to find the deuteron in a $D-$state~\cite{Lacombe:1981eg,Buck:1979ff,Zuilhof:1980ae}.
The available data for deuteron structure function are also used during the fitting process.
\subsection{The $g_2$ structure function}
Now by accessing to the $g_1$ structure function, one can calculate $g_2$ via Wandzura--Wilczek \cite{Wandzura:1977qf,Piccione:1997zh} relation as in the
following
\begin{equation}
g_{2}(x,Q^{2})=-g_{1}^{p}(x,Q^{2})+\int_{x}^{1}\frac{dy}{y}g_{1}^{p}(y,Q^{2})~.\label{eq:xg2}
\end{equation}
We are now  at the position to investigate the fits to spin dependent structure functions, as we do it in next section, to extract the PPDFs from the available data.
\section{Fitting contents in QCD analysis \label{sec666}}

\subsection {Overview of data sets}
%----------------------------------
\begin{figure}[!htb]
	%	\vspace{-3.5cm}
	\includegraphics[clip,width=0.5\textwidth]{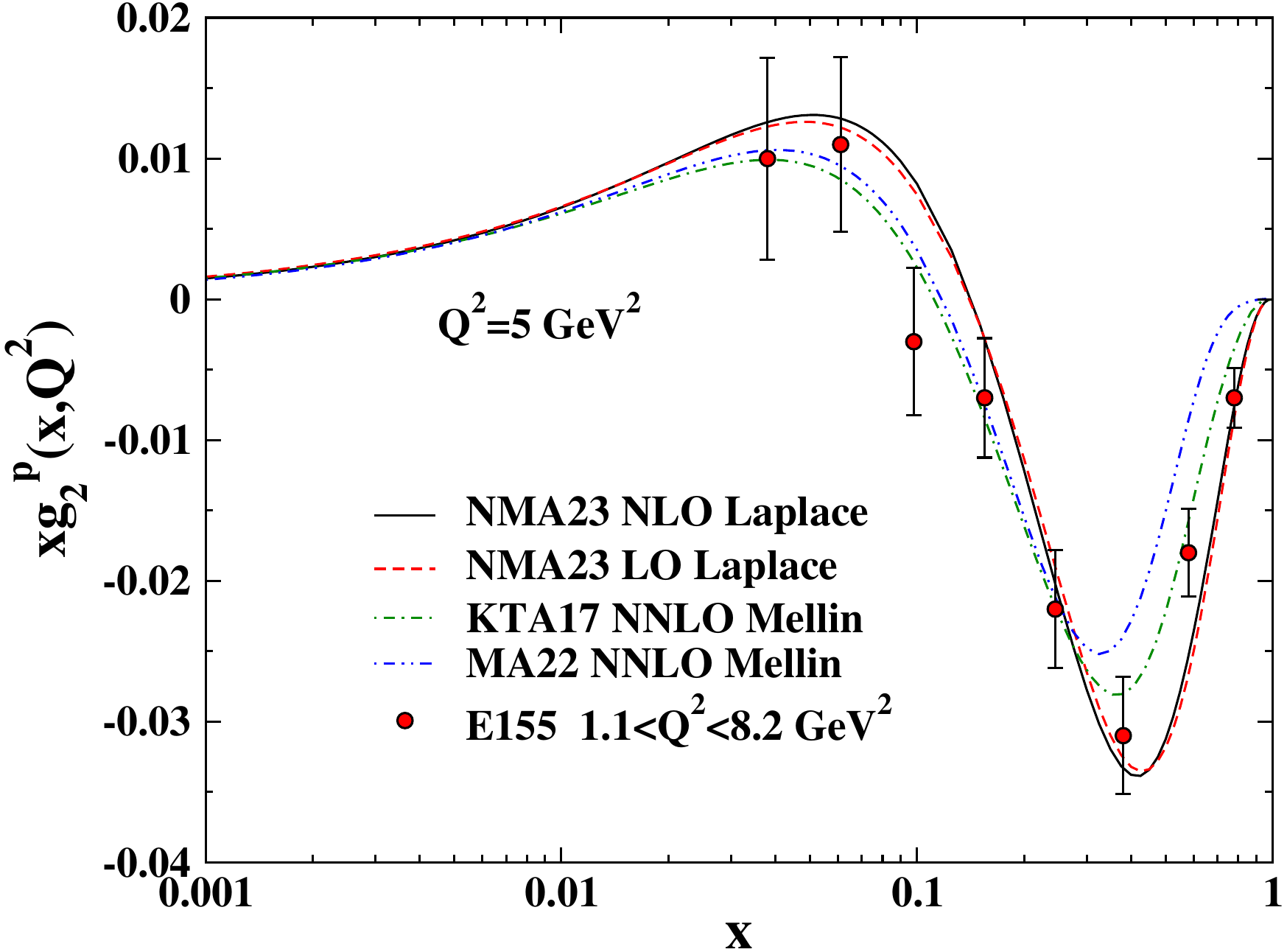}
	%\vspace{-6cm}
	\begin{center}
		\caption{{\small The polarized proton structure functions $xg_2^p$ as a function of $x$ at $Q^2$ $=5\;GeV^2$ . The results of Jacobi expansion technique in NLO (solid curve) and LO (dashed) approximations are compared with parametrization models like KTA17 (dashed-dotted)~\cite{Khanpour:2017cha}, MA22 (dashed-dotted-dotted)~\cite{Mirjalili:2022cal}.   \label{fig:xg2p}}}
	\end{center}
\end{figure}
%----------------------------------

%----------------------------------
\begin{figure}[!htb]
	%	\vspace{-3.5cm}
	\includegraphics[clip,width=0.5\textwidth]{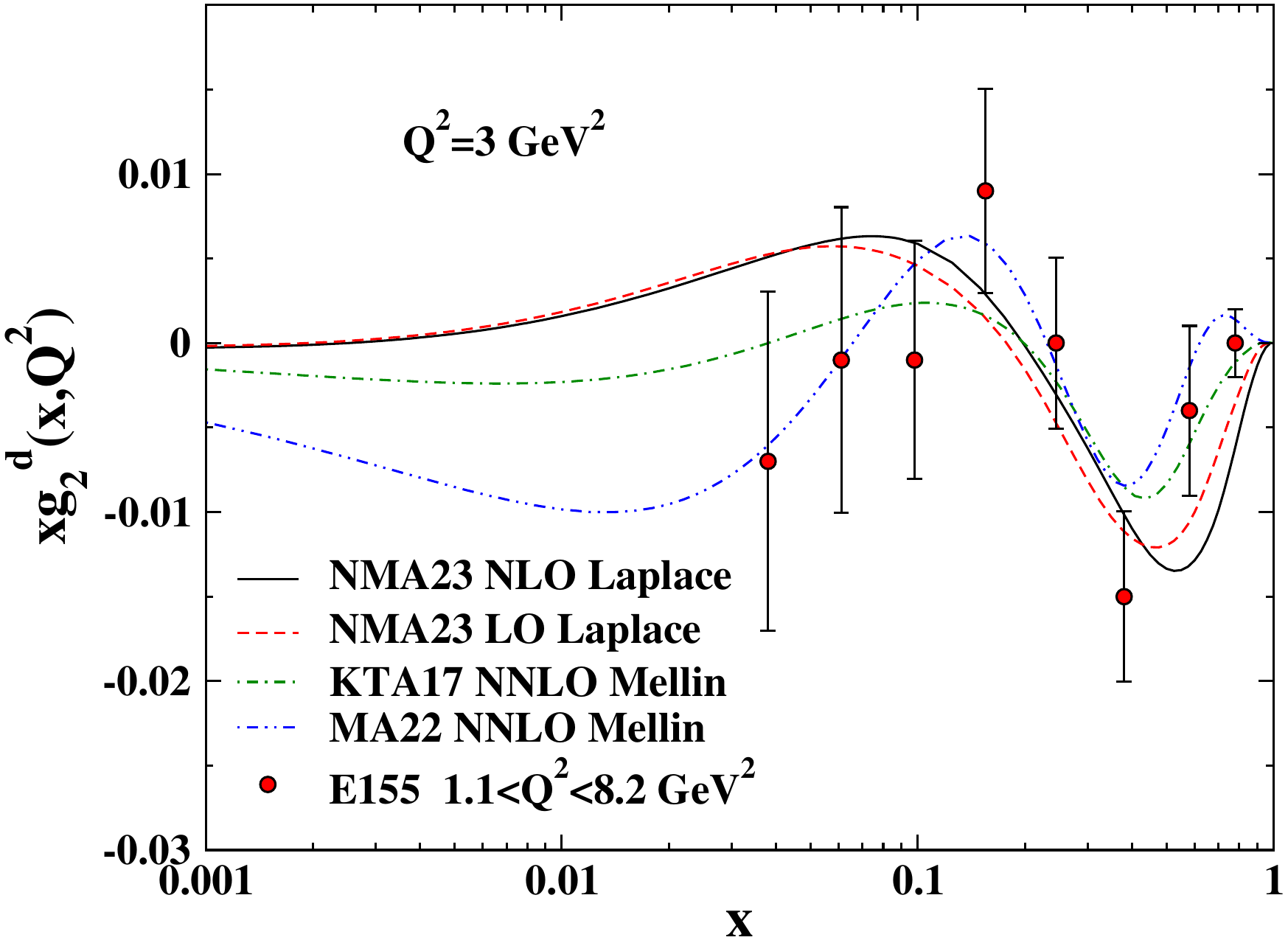}
	%\vspace{-6cm}
	\begin{center}
		\caption{{\small The spin dependent deuteron structure functions $xg_2^d$ with respect to $x$ at $Q^2$ $=3\;GeV^2$. The result of Jacobi expansion technique in NLO (solid curve) and LO (dashed) approximations are compared with parametrization models like KTA17 (dashed-dotted)~\cite{Khanpour:2017cha}, MA22 (dashed-dotted-dotted)~\cite{Mirjalili:2022cal}.   \label{fig:xg2d}}}
	\end{center}
\end{figure}
In  our recent analysis which we call it NMA23 we focus on the polarized DIS data samples. The needed DIS data for all PPDFs are coming from the experiments at electron-proton collider and also in fixed-target situation including proton, neutron and heavier targets such as deuteron.
	
Although separating quarks from antiquarks is not possible, nonetheless it is the inclusive DIS data that is included in the fit. Additionally we take into our NMA23 fitting procedure the $g_2$ structure  function. Due to the technical difficulty in operating the needed transversely polarized target, these data have been traditionally neglected before.

The data which we use in our recent analysis are up to date and including more data than we employed in our pervious analysis ~\cite{AtashbarTehrani:2013qea}. In fact we use all available data of $g_1^p$ from E143, HERMES98, SMC, EMC, E155, HERMES06, COMPASS10, COMPASS16, JLAB06 and JLAB17 experiments ~\cite{Abe:1998wq,HERM98,Adeva:1998vv,Ashman:1987hv,E155p,HERMpd,COMP1,Adolph:2015saz,Dharmawardane:2006zd,Fersch:2017qrq}, and $g_1^n$ data from HERMES98, E142, E154, HERMES06, Jlab03, Jlab04 and Jlab05~\cite{HERM98,E142n,E154n,Ackerstaff:1997ws,JLABn2003,JLABn2004,JLABn2005} and finally the data of $g_1^d$ from E143, SMC, HERMES06, E155, COMPASS05, COMPASS06 and COMPASS17~\cite{Abe:1998wq,Adeva:1998vv,HERMpd,E155d,COMP2005,COMP2006,Adolph:2016myg}.
The DIS data for $g_2^{p, n, d}$ from E143, E142, Jlab03, Jlab04, Jlab05, E155, Hermes12 and SMC~\cite{Abe:1998wq,E142n,JLABn2003,JLABn2004,JLABn2005,E155pdg2,hermes2012g2,SMCpg2}  are also included.
These data sets are listed in Table~\ref{tab:DISdata}. We also present the kinematic coverage, the number of data points for each given target, and the fitted normalization shifts ${\cal{N}}_i$ in this Table.
Our NMA23 analysis algorithm calculates the $Q^2$ evolution and extracts the polarized structure function in $x$ space using Jacobi polynomials approach. It is corresponding to the fitting programs on the market that solve the polarized DGLAP evolution equations in the Laplace space.
	
One of the important quantities used as a criteria to indicate the validation of fit process, is the chi-square ($\chi^2$) test which is assessing the goodness of fit between observed values and those expected theoretically. In next subsection we deal with about it in more details.
	\subsection{ $\chi^2$ minimization }
	%========================================================================
The goodness of fit to the data for a set of $\rm p$ independent parameters, is quantified by the $\chi_{\rm global}^2(\rm p)$ . To determine the best fit, we need to minimize the $\chi^2_{\rm global}$ function with the free unknown parameters. We perform it for PPDFs at the LO and NLO approximations that additionally include the QCD cut off parameter, $\Lambda_{\rm QCD}$ which finally yield us the polarized PDFs at Q$_0^2$ = 1.3 GeV$^2$.
		
		This function is presented as it follows:
	%--------------------------------
	%
	\begin{equation}\label{eq:chi2}
	\chi_{\rm global}^2 ({\rm p}) = \sum_{n=1}^{N_{\rm exp}} w_n \chi_n^2\,.
	\end{equation}
	%--------------------------------
	In above equation, ${w_n}$ denotes a weight factor for the $n^{\rm th}$ experiment. However this factor in principle can have different values for various data sets but since all of the experimental  data sets have identical worthiness, we take the related weight factor equal to one in our analyses \cite{Pumplin:2001ct,Paukkunen2014,Han2022}. Following that the $\chi_n^2$ in Eq.(\ref{eq:chi2}) is defined as:
	%--------------------------------
	\begin{equation}\label{eq:chi2global}
	\chi_n^2 (\rm p) = \left( \frac{1 -{\cal N}_n }{\Delta{\cal N}_n}\right)^2 + \sum_{i=1}^{N_n^{\rm data}} \left(\frac{{\cal N}_n  \, g_{(1,2), i}^{\rm Exp} - g_{(1,2), i}^{\rm Theory} (p) }{{\cal N}_n \, \Delta g_{(1,2), i}^{\rm Exp}} \right)^2\,.
	\end{equation}
	%----------------------------------
	%
	The minimization of the $\chi_{\mathrm{\rm global}}^2 (\rm p)$ function is done applying the CERN program library MINUIT~\cite{James:1994vla}.
	In the above equation, the essential contribution originates from the difference between the model and the DIS data within the statistical precision.
 In the $\chi_n^2$ function, $g^{\rm Theory}$ indicates the theoretical value for the $i^{\rm th}$ data point and $g^{\rm Exp}$, $\Delta g^{\rm Exp}$ denote the experimental measurement and the experimental uncertainty, respectively that is coming from statistical and systematic uncertainties, combined in quadrature.
		
		To do a proper fit, we need an over normalization factor for the data of experiment $n$ that is denoted by ${\cal N}_n$. An uncertainty ${\Delta{\cal N}_n}$ is attributed to this factor which should be regarded in the fit. These factors, considering the uncertainties, quoted by the experiments are used to relate different experimental data sets. They are taken as free parameters determined simultaneously with the other parameters in the fit process. In fact they are obtained in the  pre-fitting procedure and then fixed at their best values in further steps.
		Numerical results of the unknown parameters, obtained from $\chi^2$ minimization, are listed in Table.\ref{tab:fit}. Different data sets, used in the fit process, are presented in Table.\ref{tab:DISdata}.

We should remind that  the results of fitting process for PPDFs at initial $Q_0^2$ energy scale is based on Hessian approach \cite{Pumplin:2001ct} such that $\Delta \chi^2/\chi^2 =0.2 \%$  that is corresponding to bigger confidence level (C.L)  than the inconvenience choice $\Delta \chi^2 =1$ with 68$\%$ C.L \cite{deFlorian2009}. The plots  for polarized parton  densities  are depicted in  Fig.\ref{fig:partonLOQ0} and Fig.\ref{fig:partonNLOQ0} on this base.

		\subsubsection{Gluon and sea quarks}
		We find the factor $(1+c_{q}x)$ in Eq.~(\ref{eq:parm}), provides the
		flexibility to obtain a good description of the data, especially
		for the polarized valence quark distributions $\Delta u_{v},\Delta d_{v}$. Thus we will make
		use of the $c_{q}$ coefficients for the up-valence and
		down-valence quark distribution functions; in contrast, we are able to set the
		values of $c_{\bar{q}}$ and $c_{g}$ to zero,
		$(c_{\bar{q}}=c_{g}=0)$, while preserving a good fit and
		eliminating two free parameters. We find the fit improves if we use non-zero values for the $c_{u_{v}}, c_{d_{v}}$
		parameters, but as these are relatively flat directions in $\chi$-space we shall
		fix the values as detailed in Table~\ref{tab:fit}.

		Having fixed $\eta_{u_{v}},\eta_{d_{v}}$ and $b_{\bar{q}},b_{g}$  parameters in preliminary
		minimization and to take $c_{\bar{q}}=c_{g}=0$ which we referred before, we then set the
		$b_{\bar{q}},b_{g},c_{u_{v}},c_{d_{v}}$ parameters as indicated in Table~\ref{tab:fit}; this gives us a total of 9 unknown parameters, in addition to $\alpha_{s}(Q_{0}^{2})$.
\begin{table}
\begin{center}
\begin{tabular}{cccccc}
\hline
\multicolumn{6}{c}{LO} \\
\hline
					$\Delta u_{v}$ & $\eta _{u_{v}}$ & $0.928$(Fixed) & $\Delta \overline{q}$ & $%
					\eta _{\overline{q}}$ &  $-0.076068\pm 0.0017283$ \\
					& a$_{u_{v}}$ & $0.2906\pm 0.01061$ &  & a$_{\overline{q}}$ & $0.42486\pm 0.03115
					$ \\
					& b$_{u_{v}}$ & $2.0498\pm 0.010617$ &  & b$_{\overline{q}}$ & $2.7562$(Fixed) \\
					& c$_{u_{v}}$ & $16.4977$(Fixed) &  & c$_{\overline{q}}$ & $0$ \\ \hline
					$\Delta d_{v}$ & $\eta _{d_{v}}$ & $-0.342$(Fixed) & $\Delta g$ & $\eta _{g}$
					& $1.1543\pm 0.1792$ \\
					& a$_{d_{v}}$ & $0.1274\pm 0.003728$ &  & a$_{g}$ & $2.4164\pm 0.3716$ \\
					& b$_{d_{v}}$ & $1.8621\pm 0.044186$ &  & b$_{g}$ & $1.7430$(Fixed) \\
					& c$_{d_{v}}$ & $35.7909$(Fixed) &  & c$_{g}$ & $0$ \\
					\multicolumn{6}{c}{$\Lambda =0.2007\pm 0.05004GeV$} \\
                    \multicolumn{6}{c}{$\alpha(M_z^2) =0.12812\pm 0.0038$} \\
					\multicolumn{6}{c}{$\chi ^{2}/D.O.F=1171.65/957=1.224$} \\ \hline
					\multicolumn{6}{c}{NLO} \\
					\hline
					$\Delta u_{v}$ & $\eta _{u_{v}}$ & $0.928$(Fixed) & $\Delta \overline{q}$ & $%
					\eta _{\overline{q}}$ &  $-0.076272\pm 0.001742$ \\
					& a$_{u_{v}}$ & $0.33889\pm 0.01121$ &  & a$_{\overline{q}}$ & $0.4844\pm 0.029784
					$ \\
					& b$_{u_{v}}$ & $2.1075\pm 0.052585$ &  & b$_{\overline{q}}$ & $3.3948$(Fixed) \\
					& c$_{u_{v}}$ & $15.3475$(Fixed) &  & c$_{\overline{q}}$ & $0$ \\ \hline
					$\Delta d_{v}$ & $\eta _{d_{v}}$ & $-0.342$(Fixed) & $\Delta g$ & $\eta _{g}$
					& $0.2526\pm 0.05528$ \\
					& a$_{d_{v}}$ & $0.1294\pm 0.004110$ &  & a$_{g}$ & $2.0898\pm 0.4409$ \\
					& b$_{d_{v}}$ & $1.8654\pm 0.043007$ &  & b$_{g}$ & $1.0174$(Fixed) \\
					& c$_{d_{v}}$ & $41.9067$(Fixed) &  & c$_{g}$ & $0$ \\
                     \multicolumn{6}{c}{$a_1 =0.0250$(Fixed) } \\
                    \multicolumn{6}{c}{$b_1 =10.70$(Fixed) } \\
                    \multicolumn{6}{c}{$a_0 =0.2399$(Fixed) } \\
					\multicolumn{6}{c}{$\Lambda =0.2156\pm 0.04989GeV$} \\
                    \multicolumn{6}{c}{$\alpha(M_z^2) =0.115457\pm 0.00341$} \\
					\multicolumn{6}{c}{$\chi ^{2}/D.O.F=1128.98/957=1.179$} \\ \hline
				\end{tabular}
				\caption{{\small Final parameter values and their statistical errors in the
						$\overline{{\rm MS}}$--scheme at the input scale $Q_{0}^{2}=1.3$ GeV$^{2}$.}
					\label{tab:fit}}
\end{center}
\end{table}

 Now in order to validate the results of the fitting, we consider and calculate some sum rules as we do it in next section.
\section{The Sum Rules \label{sec:sec7}}
		%========================================================================
Some fundamental properties of the nucleon structure can be inspected by considering QCD sum rules like total momentum fraction carried by partons and also the total contribution of parton spin to the spin of the nucleon. In what are following by utilizing available experimental data, we analysis some important polarized sum rules.
		 \subsection{Bjorken sum rule} \label{Bjorken-sum-rule}
		 %========================================================================
Integral over the spin distributions of quarks inside the nucleon yields to the polarized Bjorken sum. It can be written in terms of multiplication of nucleon axial charge, $g_A$ (as measured in neutron $\beta$ decay) with a coefficient function, $C_{Bj}[\alpha_s(Q^2)]$.
Taking into account the corrections of higher twist (HT), this sum rule is given by ~\cite{Bjorken:1969mm}:
		 %----------------------------------
		 \begin{eqnarray}\label{eq:Bjorken-SR}
		 \Gamma_1^{\rm NS}(Q^2)&=&\Gamma_1^p(Q^2) - \Gamma_1^n(Q^2)   \nonumber \\
		 &=& \int_0^1[g_{1}^{p}(x, Q^2) - g_1^{{n}}(x, Q^2)]dx        \nonumber \\
		 &=& \frac{1}{6} ~ |g_A|~ C_{Bj}[\alpha_s(Q^2)] + \text{HT corrections}\,.\nonumber \\
		 \end{eqnarray}
		 %----------------------------------
A very precise determination on the $\alpha_s$ as strong coupling constant can be provided by Bjorken sum rule. Using $C_{Bj}[\alpha_s(Q^2)]$ expression the value of  coupling can be extracted from experimental data while the present world average value is $\alpha_{s}(M_{Z}^{2}) = 0.1179\pm 8.5\times10^{-6}$~\cite{Zyla:2020zbs}. At 4-loop corrections of perturbative QCD (pQCD) this function has been calculated in both massless \cite{Baikov:2010je} and massive cases \cite{Blumlein:2016xcy}. Due to ambiguities from small-$x$ extrapolation, determining $\alpha_s$ from the Bjorken sum rule is suffering ~\cite{Altarelli:1998nb}. Nevertheless in our computations the numerical value for coupling constant at $Z$-boson  mass scale can be found during the fitting process to find the unknown parameters of polarized parton densities at initial energy scale $Q_0$. Outputted results for coupling constant at LO and NLO analysis are presented in Table.~\ref{tab:fit} which are in good agreement with reported world average value of this quantity.

In Table~\ref{tab:Bjorken} we list our results for the Bjorken sum rule. Experimental measurements such as \text{E143}~\cite{Abe:1998wq}, \text{SMC}~\cite{SMCpg2}, \text{HERMES06}~\cite{HERMpd} and \text{COMPASS16}~\cite{Adolph:2015saz} are added to this table. An adequate consistency can be seen between them.

		 %
		 %%%%%%%%%%%%%%
		 % TABLE 7
		 %%%%%%%%%%%%%%
		 %
		 \begin{table*}[!htb]
		 	\caption{\label{tab:Bjorken} Our computed LO and NLO results for the Bjorken sum rule, $\Gamma_1^{NS}$, in comparison with world data from
		 		\text{E143}~\cite{Abe:1998wq}, \text{SMC}~\cite{SMCpg2}, \text{HERMES06}~\cite{HERMpd} and \text{COMPASS16}~\cite{Adolph:2015saz}.
		 		Only HERMES06~\cite{HERMpd} results are not extrapolated in full $x$ range (measured in region $0.021 \leq x \leq 0.9$). }
		 	\begin{ruledtabular}
		 		\begin{tabular}{lcccccc}
		 			& \textbf{E143}~\cite{Abe:1998wq}   & \textbf{SMC}~\cite{SMCpg2}  & \textbf{HERMES06}~\cite{HERMpd}  &  \textbf{COMPASS16}~\cite{Adolph:2015saz} & \textbf{LO} & \textbf{NLO}\\ % \tabularnewline\\
		 			& $Q^2=5$ GeV$^2$& $Q^2=5$ GeV$^2$& $Q^2=5$ GeV$^2$& $Q^2=3$ GeV$^2$&$Q^2=5$ GeV$^2$  &$Q^2=5$ GeV$^2$     \\     \hline    \hline\tabularnewline
		 			$\Gamma^{\rm NS}_1$   & $0.164 \pm 0.021$ & $ 0.181\pm 0.035 $  & $0.148 \pm 0.017$ & $0.181\pm 0.008$& $0.15632\pm0.0062$ & $0.15350\pm0.00081$    \\
		 		\end{tabular}
		 	\end{ruledtabular}
		 \end{table*}
		 %%%%%%%%%%%%%%
%========================================================================
\subsection{Proton helicity sum rule}
%========================================================================
In order to complete our knowledge in the field of nuclear physics an extrapolation of proton spin among its constituents can be done and consequently new sum rule as  proton helicity sum rule is achieved \cite{Leader:2016sli}. Considering this sum rule, by a precise extraction of PPDFs, one can obtain an accurate picture of the quark and gluon helicity densities.

Since each constituent of a nucleon is carrying part of nucleon spin, the total spin of nucleon can be written as:
%----------------------------------
\begin{equation}\label{eq:spinsumrule}
\frac{1}{2} = \frac{1}{2} \Delta \Sigma(Q^2) + \Delta {\mathrm G}(Q^2) + {\mathrm L}(Q^2).
\end{equation}
%----------------------------------
In this equation $\Delta \Sigma(Q^2)=\sum_{i}\int_{0}^{1}dx~(\Delta q(x,Q^2)+\Delta \bar{q}(x,Q^2))$ represents spin contribution of the singlet flavour, $\Delta {\rm G(Q^2)}=\int_{0}^{1}dx~\Delta g(x,Q^2)$  denotes the gluon spin contribution and finally ${\mathrm L}(Q^2)$ is interpreted as the total contribution from quark and gluon orbital angular momentum. In  Eq.(\ref{eq:spinsumrule}) each term depends on $Q^2$ but the sum of them does not. The measuring processes of them can not be done easily and it is  beyond the scope of this paper to describe their measurement methods.

Numerical values of first moments of the singlet-quark and gluon at Q$^2$=10 GeV$^2$ are listed in  Table~\ref{tab:firstmomentQ10}.
Our results at both truncated and full $x$ region are compared to those from the  \text{NNPDFpol1.1}~\cite{Nocera:2012hx} and \text{DSSV14}~\cite{deFlorian:2014yva}.
	
As can be seen from Table~\ref{tab:firstmomentQ10} for the $\Delta \Sigma$, our NMA23 results are consistent, within uncertainty, with those of other groups. This is occurred because in semileptonic decays the first moment of polarized densities are mainly fixed.
Very different values are reported by various groups when the gluon contribution is considered. Due to their large uncertainty we are avoided to get a stiffen result for the full first moment of gluon.

% TABLE
%%%%%%%%%%%%%%
%
\begin{table*}[!htb]
	\caption{\label{tab:firstmomentQ10} Results for the full and truncated first moments of the polarized singlet-quark $\Delta \Sigma(Q^2)=\sum_{i}\int_0^1 dx [\Delta q_i(x) + \Delta \bar q_i (x)]$ and gluon distributions at the scale Q$^2$=10 GeV$^2$ in the $\overline{{\rm MS}}$--scheme. The recent polarized global analysis of \text{NNPDFpol1.1}~\cite{Nocera:2012hx} and \text{DSSV14}~\cite{deFlorian:2014yva} are also presented. }
	\begin{ruledtabular}
		\begin{tabular}{lcccc}
			& \textbf{DSSV14}~\cite{deFlorian:2014yva}  &  \textbf{NNPDFpol1.1}~\cite{Nocera:2012hx}&  \textbf{LO}  &  \textbf{NLO}   \tabularnewline
			Full $x$ region $[0,1]$ &    & & & \\
			\hline \hline \tabularnewline
			$\Delta \Sigma {\rm(Q^2)}$&$0.291799$  &$+0.18\pm 0.21$       &$0.12959\pm0.00922$  &$0.148546\pm0.0194$  \\
			$\Delta {\rm G(Q^2)}$   & $0.37109$ & $0.03\pm 3.24$     &$1.4693\pm1.049$ &$0.8626\pm0.3054$ \\ \hline  \hline  \tabularnewline
			Truncated $x$ region [$10^{-3},1$] &  &   & & \\     \hline \hline \tabularnewline
			$\Delta \Sigma {\rm(Q^2)}$  & $0.36645$&$+0.25\pm0.10$    &$0.03644\pm0.0406$ &$0.04848\pm0.0187$\\
			$\Delta {\rm G(Q^2)}$   &$0.3636$  & $0.49\pm 0.75$  &$1.3848\pm0.9829$ & $0.7827\pm0.2854$
		\end{tabular}
	\end{ruledtabular}
\end{table*}
%%%%%%%%%%%%%%
%
% TABLE
%%%%%%%%%%%%%%

%%%%%%%%%%%%%%
%
\begin{table*}[!htb]
	\caption{\label{tab:twist3} $d_2$ moments of the proton, neutron and deuteron polarized structure functions from the \text{SLAC E155x}~\cite{E155pdg2},
		\text{E01-012}~\cite{Solvignon:2013yun}, \text{E06-014}~\cite{Flay:2016wie}, \text{Lattice QCD}~\cite{Gockeler:2005vw}, \text{CM bag model}~\cite{Song:1996ea}, \text{JAM15}~\cite{Sato:2016tuz}, \text{JAM13}~\cite{Jimenez-Delgado:2013boa} compared with LO and NLO results.
	}
	\begin{ruledtabular}
		\begin{tabular}{lccccc}
			& Ref. &  \textbf{$Q^2$ [GeV$^{2}$]} &  \textbf{$10^2d^{p}_2$}  & \textbf{$10^5d^{n}_2$}  & \textbf{$10^3d^d_2$} \tabularnewline  \hline  \hline  \tabularnewline
			LO      &&$5$&$0.2994\pm0.00035$         &    $127.68\pm11.0073$     & $1.2534\pm0.02719$       \\
			NLO      & &$5$&$0.2855\pm0.00196$         &   $23.4919\pm0.6654$     & $1.3464\pm0.0743$       \\
			\text{E06-014} & \cite{Flay:2016wie}& 3.21 & &$-421.0 \pm 79.0 \pm 82.0 \pm 8.0$ & -\\
			\text{E06-014} & \cite{Flay:2016wie}& 4.32 & &$-35.0 \pm 83.0 \pm 69.0 \pm 7.0$  & -\\
			\text{E01-012} & \cite{Solvignon:2013yun} &3 & - & $-117 \pm 88 \pm 138$ &  -  \\
			\text{E155x}   & \cite{E155pdg2} & $5$ & $0.32\pm 0.17$ & $790\pm 480$ &     -                         \\
			\text{E143}    & \cite{Abe:1998wq}& $5$                 & $0.58\pm 0.50$ & $500\pm 2100$ & $5.1\pm 9.2$ \\
			\text{Lattice QCD}  &\cite{Gockeler:2005vw} & 5 &  0.4(5) & -100(-300) & - \\
			\text{CM bag model} &\cite{Song:1996ea} & $5$  & $1.74$          & $-253$       & $6.79 $         \\
			\text{JAM15}        &\cite{Sato:2016tuz} & $1$  & $0.5\pm 0.2$          & $-100 \pm 100$       & -        \\
			\text{JAM13}        &\cite{Jimenez-Delgado:2013boa} & $5$  & $1.1 \pm 0.2$          & $200 \pm 300$       &  -         \\
			
		\end{tabular}
	\end{ruledtabular}
\end{table*}
%%%%%%%%%%%%%%
%
%%%%%%%%%%%%%%
% TABLE BC
%%%%%%%%%%%%%%
%
\begin{table*}[!htb]
	\caption{ \label{tab:BC} The result of BC sum rule for $\Gamma_2^p$, $\Gamma_2^d$ and $\Gamma_2^n$ in comparison with world data from \text{E143}~\cite{Abe:1998wq}, \text{E155}~\cite{E155pdg2}, \text{HERMES2012}~\cite{hermes2012g2}, \text{RSS}~\cite{Slifer:2008xu}, \text{E01012}~\cite{Solvignon:2013yun}. }
	\begin{ruledtabular}
		\begin{tabular}{lccccccc}
			
			& \textbf{E143}~\cite{Abe:1998wq}  & \textbf{E155}~\cite{E155pdg2}  & \textbf{HERMES2012}~\cite{hermes2012g2}& \textbf{RSS}~\cite{Slifer:2008xu} & \textbf{E01012}~\cite{Solvignon:2013yun} & \textbf{LO}& \textbf{NLO}   \\
			& $0.03 \le x \le 1 $  & $0.02 \le x \le 0.8 $   & $0.023 \le x \le 0.9$  &$0.316 < x < 0.823 $&$0 \le x \le 1 $& $0.03 \le x \le 1 $& $0.03 \le x \le 1 $\\
			& $Q^2=5$ GeV$^2$& $Q^2=5$ GeV$^2$& $Q^2=5$ GeV$^2$& $Q^2=1.28$ GeV$^2$&$Q^2=3$ GeV$^2$&$Q^2=5$ GeV$^2$&$Q^2=5$ GeV$^2$   \\
			\hline  \hline
			$\Gamma_2^p$  & $-0.014 \pm 0.028$ &$-0.044 \pm 0.008$ & $0.006 \pm 0.029$ & $-0.0006 \pm 0.0022$&...&  $ -0.01911 \pm 0.0199$ &  $ -0.01929 \pm 0.00038$  \\
			$\Gamma_2^d$  & $-0.034 \pm 0.082$ &$-0.008 \pm 0.012$ &-  &$-0.0090 \pm 0.0026$&...& $ -0.001687 \pm 0.000016$ & $ -0.0028986 \pm0.00053$   \\
			$\Gamma_2^n$ &-&-&-&$-0.0092 \pm 0.0035$&$0.00015 \pm 0.00113$& $ ~0.007824 \pm0.00056$& $ ~0.0034500 \pm 0.000016$  \\
		\end{tabular}
	\end{ruledtabular}
\end{table*}
%%%%%%%%%%%%%%
The proton spin sum rule can be finally calculated, considering the extracted values which are listed in Table~\ref{tab:firstmomentQ10}.
Accordingly numerical value of quark and gluon orbital angular momentum, attributed to the spin of the proton, is obtained as:

%----------------------------------
\begin{equation}
{\mathrm L}(Q^2=10~\rm GeV^2) = %\frac{1}{2} - \frac{1}{2} (0.124 \pm 0.029) -(0.167 \pm 0.036) =
-0.436873 \pm 0.334587 \,.
\end{equation}
%----------------------------------
The contribution of total orbital angular momentum to the spin of the proton can not be determined tightly and it is due to the large uncertainty which is mostly coming out from the gluons. By improving the current level of experimental accuracy, precise determination of each individual contribution to the nucleon spin can be obtained.
\subsection{The twist-3 reduced matrix element $d_2$}
%========================================================================%
Twist-3 reduced matrix element, denoted by $d_2$, is not considered as a sum rule but to investigate the higher twist effect, the numerical evaluation of this quantity is important. One can find in \cite{Blumlein:2010rn} the detailed analyses of higher twist, related to the $g_1$ polarized structure function. Through the moments of $g_1$ and $g_2$ structure functions, considering the operator product expansion (OPE) theorem \cite{ope}, the effect of quark-gluon correlations can be studied. These considerations for the moments will conclude the following definition for $d_2(Q^2)$ as reduced matrix element:
%----------------------------------
\begin{eqnarray}
d_2(Q^2) &=& 3 \int_0^1 x^2 \bar{g_2}(x,Q^2)~dx       \nonumber   \\
&=& \int_0^1 x^2 [3 g_2(x, Q^2) + 2g_1(x, Q^2)]~ dx .
\label{d2}
\end{eqnarray}
%----------------------------------
In above equation we have $\bar g_2=g_2-g_{2} ^{WW}$ where $g_{2} ^{WW}$, corresponding to Eq.(\ref{eq:xg2}), is given by Wandzura and Wilczek (WW). The deviation of $g_2$ from $g_{2}^{\tau_2}$ which is polarized structure function at leading twist order can be measured, using the $d_2(Q^2)$ as the twist-3 reduced matrix element of spin dependent operators in nucleon.
This matrix element because of the $x^2$ weighting factor in Eq.(\ref{d2}) has remarkable sensitive to the behaviour of $\bar{g_2}$ at large-$x$ values.
By extracting the $d_2$ term, valuable intuition about the size of the multi-parton correlation terms can be achieved which denotes the importance of this quantity.

Having non-zero value for $d_2$  reveals us the importance of higher twist terms in QCD analyses. To improve model prediction, more information on the higher twist operators are required and this can be done by precise measurement of $d_2$ term.
Our results for $d_2$, compared with experimental values and also some theoretical predictions, are presented in Table~\ref{tab:twist3}.
%========================================================================
\subsection{Burkhardt-Cottingham  sum rule} \label{BC-sum-rule}
%========================================================================
The zeroth moment of $g_2$ structure function, considering dispersion relations for virtual Compton scattering at all $Q^2$ values, is predicted to get zero value and consequently Burkhardt and Cottingham (BC) sum rule is obtained such as ~\cite{Burkhardt:1970ti}:
%----------------------------------
\begin{equation}
\Gamma_2 = \int_0^1 dx \, g_2(x, Q^2) = 0~.
\end{equation}
%----------------------------------
BC sum rule is insignificant result, arising out from WW relation which is given by Eq.(\ref{eq:xg2}). In light cone expansion, one can not obtained the zeroth moment of structure function and hence local operator product expansion \cite{Blumlein:1996vs} can not describe this moment. This sum rule can be used even if the structure function involves target mass correction \cite{Blumlein:1998nv}. Finally it should be said that the presence of HT contribution is denoting to violation of the BC sum rule ~\cite{hermes2012g2}.

In Table.~\ref{tab:BC} we list our results for $\Gamma_2$ at LO and NLO approximations where data from \text{E143}~\cite{Abe:1998wq}, \text{E155}~\cite{E155pdg2}, \text{HERMES2012}~\cite{hermes2012g2}, \text{RSS}~\cite{Slifer:2008xu}, \text{E01012}~\cite{Solvignon:2013yun} groups for proton, deuteron and neutron are also added there. The behaviour of $g_2$ structure function at low-$x$ values has not yet measured accurately but it has significant effect on any possible conclusion which we get.
\section{Conclusion \label{sec:sec8}}
For about three decades, studies of the internal spin structure of the proton have advanced
steadily using the technology of polarized beams and polarized targets. The demand for higher-energy experiments to access the deepest regions inside the proton and to extract the theoretically cleanest results continues.

Polarized deep inelastic scattering remains one of the cleanest tests for studying the
internal spin structure of the proton and neutron. Pioneering experiments at SLAC, scattering polarized electrons off polarized protons, helped to establish the quark structure of the proton with the observation of large spin-dependent asymmetries.
An exciting follow-up experiment (CERN EMC) at higher energies uncovered a violation of a quark parton model sum rule, implying that the quarks accounted for only a small fraction of the proton spin and giving birth to the proton-spin crisis. Since about three decades ago a large sample of data from polarized fixed-target experiments at SLAC, CERN,
and DESY have resulted in a substantial perturbative-QCD analysis of the nucleon
spin structure.

Determining the nucleon spin structure functions $g_1(x,Q^2)$ and $g_2(x,Q^2)$ and their moments is the main goal of our present NMA23 analysis. They are essential to test some QCD sum rules .We provided a unified and consistent PPDF through an achievement, containing an appropriate description of the fitted data. Within the known very large uncertainties arising from the lack of constraining data, our helicity distributions are in good consistency with other extractions. We studied Bjorken sum rule, proton helicity  and  Burkhardt-Cottinghan sum rules. Our results for the reduced matrix element d2 at the NLO approximation have also been presented. To investigate them precisely, more accurate data are needed and in this work we considered all the recent and available data, relating to polarized targets which are listed in Table.\ref{tab:DISdata}.

Using Jacobi polynomial technique a fit to the polarized lepton DIS data on nucleon have been presented at the NLO approximation.
We found good agreement with the experimental data and our results have been in correspond with determinations from some
parametrization models. In general we could demonstrate that an acceptable progressive has been achieved to describe the
spin structure of the nucleon.

The available data which we use in our recent analysis are up to date and including more data than we employed in our pervious analysis \cite{AtashbarTehrani:2013qea}. These data sets are summarized in Table \ref{tab:DISdata}. The kinematic
coverage, the number of data points for each given target, and the fitted normalization shifts $N_i$ also presented in this
table. Our NMA23 analysis algorithm computed the $Q^2$ evolution and extracted the spin structure functions in $x$ space using Jacobi polynomials approach. It corresponded to the fitting programs of other groups which solve the DGLAP evolution equations
in the Mellin space.
Results for PPDFs at different energy scales and $g_1^{p,n}$ structure functions together with deuteron nucleus have been presented in Fig.\ref{fig:partonLOQ0} to Fig.\ref{fig:xg2d} which confirm the validity of computations during the fitting process to all updated and recent related data.\\

The current analysis can be extended to include transverse polarized targets, using the Jacobbi polynomial expansions in Laplace $s$-space or other polynomial expansion which can be done as our further research task.
%%%%%%%%%%%%%%%%%%%%%%%%%%%%%%%%%%%%%%%%%%%%%%%%%%%%%%%%%
\section*{Acknowledgments}
H.~N is indebted to Shaid Bahonar university of Kerman and A.~M acknowledges Yazd university for the provided facilities to do this project. S. A. T. is grateful to the School of Particles and Accelerators, Institute for Research in Fundamental Sciences (IPM) to make the required support to do this project.

%%%%%%%%%%%%%%%%%%%%%%%%%%%%%%%%

\end{document}